\documentclass[aps,pra,10pt,twocolumn,superscriptaddress,floatfix,nofootinbib,showpacs,longbibliography]{revtex4-1}

\usepackage{makecell}
\usepackage{array}
\newcolumntype{C}[1]{>{\centering\arraybackslash}m{#1}}
\usepackage{booktabs,adjustbox}
\usepackage{float}
\expandafter\let\csname equation*\endcsname\relax

\expandafter\let\csname endequation*\endcsname\relax
\usepackage{amsmath}
\usepackage[normalem]{ulem}
\usepackage[utf8]{inputenc}  
\usepackage[T1]{fontenc}     
\usepackage[table]{xcolor}
\usepackage{color,soul}
\usepackage[british]{babel}  
\usepackage[colorlinks=true, citecolor=blue, urlcolor=blue]{hyperref}  
\usepackage{graphicx} 
\usepackage[babel]{microtype}  
\usepackage{dsfont,amsthm}
\usepackage{amsmath,amssymb,amsthm,bm,amsfonts,mathrsfs,bbm,dsfont} 
\usepackage{xspace}  
\usepackage{pgf,tikz}
\usepackage{xcolor}
\usepackage{multirow}
\usepackage{array}
\usepackage{bigstrut}
\usepackage{braket}
\usepackage{color}
\usepackage{multirow}
\usepackage{mathtools}
\usepackage{float}
\usepackage{blkarray}
\usepackage[caption = false]{subfig}
\usepackage{xcolor,colortbl}
\usepackage{color}
\usepackage{rotating}
\usepackage{tikz}
\usepackage{tabularx}
\usepackage{comment}
\usepackage[shortlabels]{enumitem}
\usepackage[normalem]{ulem}

\newcommand{\be}{\begin{equation}}
\newcommand{\ee}{\end{equation}}
\newcommand{\ba}{\begin{eqnarray}}
\newcommand{\ea}{\end{eqnarray}}

\newtheorem{theorem}{Theorem}
\newtheorem{corollary}{Corollary}
\newtheorem{definition}{Definition}
\newtheorem{proposition}{Proposition}

\newtheorem{lemma}{Lemma}

\definecolor{darkorange}{rgb}{1.0, 0.55, 0.0}

\def\>{\rangle}
\def\<{\langle}

\newcommand{\map}[1]{\mathcal{#1}}

\usepackage{centernot}
\usepackage{subfig}

\begin{document}

\title{Activation of thermal states by coherently controlled thermalization processes}

\author{Kyrylo Simonov}
\affiliation{Fakult\"{a}t f\"{u}r Mathematik, Universit\"{a}t Wien, Oskar-Morgenstern-Platz 1, 1090 Vienna, Austria}
\author{Saptarshi Roy}
\affiliation{QICI Quantum Information and Computation Initiative, Department of Computer Science, The University of Hong Kong, Pokfulam Road, Hong Kong}
\author{Tamal Guha}
\affiliation{QICI Quantum Information and Computation Initiative, Department of Computer Science, The University of Hong Kong, Pokfulam Road, Hong Kong}
\author{Zolt\'an Zimbor\'as}
\affiliation{Algorithmic Ltd, Kanavakatu 3C‌, Helsinki, 00160, Finland}
\affiliation{QTF Centre of Excellence, Department of Physics, University of Helsinki, Helsinki 00100, Finland}
\affiliation{Wigner Research Centre for Physics, H-1525, P.O.Box 49, Budapest, Hungary}
\affiliation{Faculty of Informatics, E\"otv\"os Loránd University, Pázmány Péter sétány 1/C, 1117 Budapest, Hungary}
\author{Giulio Chiribella} 
\affiliation{QICI Quantum Information and Computation Initiative, Department of Computer Science, The University of Hong Kong, Pokfulam Road, Hong Kong}
\affiliation{Quantum Group, Department of Computer Science, University of Oxford, Wolfson Building, Parks Road, Oxford, OX1 3QD, United Kingdom}
\affiliation{Perimeter Institute for Theoretical Physics, 31 Caroline Street North, Waterloo, Ontario, Canada} 

\date{\today}


\begin{abstract}
Thermalization processes degrade the states of any working medium, turning any initial state into a  passive state from which no work can be extracted.  Recently, it has been shown that this degradation can be avoided if two identical thermalization processes take place in coherently controlled order, in a scenario known as the quantum SWITCH.  In some situations,  control over the order even  enables work extraction when  the medium was initially in a passive state.  This activation phenomenon, however, is subject to a  limitation:  to extract non-zero work, the initial temperature  of the medium should be less than half of the temperature of the reservoirs. Here  we analyze this limitation, showing that  it still holds true even when the medium interacts with $N\ge 2$ reservoirs  in a coherently-controlled order. Then, we show that the limitation  can be  lifted when the medium and the control systems are  initially correlated. In particular, when the medium and control are entangled, work extraction becomes possible for every initial value of the local temperature of the medium.  
\end{abstract}

\maketitle

%
%
%
%
%
\section{Introduction}

Thermalization  is the prototype of an irreversible thermodynamical process.   When interacting with a thermal bath of infinite heat capacity, macroscopic systems typically lose  their initial state,  eventually converging to a thermal distribution with the same temperature as the bath~\cite{Zemansky1968}. Thermalization phenomena are not limited to the macroscopic scale, but can also take place in microscopic systems, whose dynamics are governed by the laws of quantum mechanics~\cite{Riera2012, Gogolin2016}.  In the quantum regime, the overall  effect of a complete thermalization process is described by a quantum channel (completely positive trace-preserving map)~\cite{Heinosaari2011, Holevo2013}: specifically, it is described by a thermalization channel,  which transforms any initial state of the system into a Gibbs  state at the same temperature  of the reservoir. Since Gibbs states have minimal free energy~\cite{Huang_1987}, thermalization can be interpreted as a  complete degradation of thermodynamic resources. This degradation also affects the \textit{ergotropy}~\cite{Allahverdyan2004}, namely the maximum amount of work that can be extracted  from the system  through a unitary evolution.    Indeed, Gibbs states are known to be \textit{completely passive}~\cite{Lenard1978,Pusz1978}, meaning that no work can be extracted unitarily from them, even if infinitely many copies are available (see also~\cite{Skrzypczyk2015, Guha2020a} for a discussion from the quantum information perspective). 

Recently, it has been observed that the degradation of resources due to thermalization can be partially avoided if two thermalization processes are forced to act  in an indefinite order, using a setup known as the quantum SWITCH~\cite{chiribella2009beyond, Chiribella2013}. In the quantum SWITCH, the order of application of two processes  on a target system   is coherently controlled by the state of another quantum system, which in general can be put in a coherent  superposition of states triggering different orders.  Superposition states of the control system then result in an indefinite order of application of the two processes, and this indefiniteness has been found to  benefit quantum computations \cite{chiribella2009beyond, Chiribella2013, Chiribella2012, Araujo2014, Procopio2015, Araujo2017, Bisio2019, Taddei2021, Renner2022, Liu2023_Deutsch, Simonov2023_Comp, Apadula2024} and many information processing tasks, including channel discrimination~\cite{Chiribella2012, Bavaresco2021, Abbott2020, Goswami2018, Guerin2019}, quantum communication complexity~\cite{Guerin2016, wei2019experimental}, quantum metrology~\cite{Mukhopadhyay2018, Zhao2020, ChapeauBlondeau2021, Liu2023, Goldberg2023_Metrology}, and communication capacity enhancement~\cite{Ebler2018, Goswami2020, Guo2020, Caleffi2020, Bhattacharya2021, Chiribella2021L} as well as estimation of quantum resources \cite{Gao2022}. In particular, the works on quantum communication showed that  measurements on the control system can induce  cleaner evolutions on the target system: for example, two completely depolarizing channels applied in an indefinite order can give rise to non-depolarizing evolutions conditionally on the outcomes of measurements on the control~\cite{Ebler2018, Chiribella2021L}.  This idea  has found applications in the thermodynamic setting, where  the quantum SWITCH of two thermalization processes induces  non-thermal evolutions of the target system, benefiting   quantum cooling~\cite{Felce2020, Nie2022, Dieguez2023, Goldberg2023}, quantum battery charging~\cite{Chen2021, Zhu2023}, and  work extraction~\cite{Guha2020, Simonov2022, Francica2022Games}.  Notably, thermalization processes happening in an indefinite order can  even generate thermodynamic resources, transforming thermal states into states with non-zero free energy~\cite{Guha2020} and non-zero ergotropy ~\cite{Simonov2022}.   
 
The ability to extract non-zero work with thermal states and thermalization channels in an indefinite order, however, is subject to a strict limitation:  in order to produce output states with non-zero ergotropy, the target system must be initially cooler than the reservoirs used in the thermalization processes. Specifically, the initial temperature of the target system, denoted by $T_{\rm in}$, must satisfy the bound
\begin{align}\label{tbound}
T_{\rm in}  < \frac{T_{\rm bath} }2 \, ,
\end{align}
where $T_{\rm bath}$ is the temperature of the reservoirs~\cite{Simonov2022}.  When the condition ~(\ref{tbound}) is not satisfied, the quantum SWITCH outputs passive states, from which no work can be extracted unitarily.  

In this paper, we investigate the origin of the temperature bound ~(\ref{tbound})  and show how the bound can be overcome.  First, we consider a setup involving $N\ge 2$ thermalization processes taking place in an order determined by a control system.   In this setting, we show that increasing $N$ leads to larger work extraction when the  temperature bound ~(\ref{tbound})    is satisfied.    However, work extraction remains impossible  when the temperature bound~(\ref{tbound}) is violated. We then consider a scenario where the target system is initially correlated with the control system. Following~\cite{Llobet2015, Francica2017, Alimuddin2019},  we require that the target system be locally in a thermal state, while the control and the target are allowed to share classical or quantum correlations.  In this setting we show that the presence of classical correlations between the work medium and the system controlling the order allow to partially  break the temperature bound ~(\ref{tbound}), while quantum entanglement enables   work extraction for every value of the local  temperature of the target system.   

Our results identify the lack of correlations between target and control as the origin of the temperature bound~(\ref{tbound}).  Viewed together with recent results on quantum communication~\cite{Guha2023}, they highlight the value of quantum correlations between target and control as a new resource for quantum protocols boosted by coherent control over the causal order.

The remainder of the paper is organized as follows. In Section~\ref{sec:framework}, we introduce the necessary framework and recall the concepts of ergotropy and thermal maps, and the quantum SWITCH. In Section~\ref{sec:switch}, we discuss action of the quantum $N$-SWITCH on identical thermal maps and recover the threshold temperature bound $T_C$ for the activation of ergotropy. In Section~\ref{sec:corr}, we generalize the quantum SWITCH setup by allowing correlations between the input thermal state and the control qubit and exhibit the possible configuration to overcome the temperature bound. Last but not least, in Section~\ref{sec:conclusions}, we draw the conclusions and possible future directions as well as applications of the results.

\section{Framework}\label{sec:framework}
\subsection{Ergotropy of a quantum state}
A model of work extraction in the quantum domain involves  a quantum system $S$ with  Hamiltonian $H_S = \sum_k  \epsilon_k |\epsilon_k\rangle \langle \epsilon_k |$, used as a working medium.  The internal energy of the system can be modified by coupling the system to a macroscopic source, such as a classical control field, mathematically represented by an external  time-dependent potential $V_S(t)$~\cite{Allahverdyan2004}. In most applications, the potential is gradually switched on, starting from zero at some time $t_0$, and later is gradually switched off, reaching zero again at some later time $t_1> t_0$. During this period of time, the system undergoes a unitary evolution governed by the effective Hamiltonian $H_S'(t) = H_S + V_S(t)$ with $V_S(t)  =0$ for $t \le t_0$ and for $t\ge t_1$.

The average change of internal energy due to the action of the potential can then be quantified as  $\<\Delta H_S \> =  \operatorname{Tr} [ H_S  \rho_0 ] - \operatorname{Tr}[H_S  \rho_1]$ where $\rho_i$ is the system's state at time $t_i$, for $i\in  \{0,1\}$.  The average change of internal energy can be interpreted as a measure of the amount of work the system can perform under the driving of the external potential $V_S(t)$, under the assumption that all the energy drawn from the system is converted into work.    To maximize the work, the potential should be engineered so that the final state $\rho_1$ has the smallest possible energy. Since the possible potentials $V_S(t)$ generate all possible  overall unitary transformations from time $t_0$ to time $t_1$,  the maximum extractable work can be written as  \begin{align}\label{extractable}
  W(\rho_0) =  \operatorname{Tr} [ H_S  \rho_0 ]    -  \min_U  \,  \operatorname{Tr} [  H_S \,  U  \rho_0  U^\dag]  \, ,   
\end{align}
where the minimum runs over all possible unitary transformations. This leads to the following definition.
\begin{definition}[Ergotropy]
    Given a work storage $S$ in state $\rho$, the maximal work that can be extracted from it via unitary cycles defines ergotropy of $\rho$:
    \begin{equation}\label{eq:ergotropy}
    W(\rho) =  \max_U \operatorname{Tr} [ H_S  (\rho  -   U  \rho  U^\dag)],   
\end{equation}
where $H_S$ is the Hamiltonian of $S$.
\end{definition}

\subsubsection{Ergotropy and the role of quantum coherence}

Different variants of the notion of ergotropy can be obtained by restricting the optimization in Eq.~(\ref{extractable}) to certain subsets of unitary transformations.  One such variant is the {\em incoherent ergotropy}, defined as the maximum  work extractable by incoherent unitaries~\cite{Francica2020}, as highlighted in the following Definition.
\begin{definition}[Incoherent ergotropy]
    Given a work storage $S$ in state $\rho$, the maximal work that can be extracted from it via unitary transformations of the form
    \begin{equation}
        U  =  \sum_{k=0}^{d-1}   e^{ i \theta_k} \,  |\varepsilon_{\pi(k)}  \>\< \varepsilon_k|,
    \end{equation}
    where  $\theta_k  \in [ 0, 2\pi)$ is an arbitrary phase for every $k \in  \{0,1,\dots,  d-1\}$, and  $\pi$ is a permutation, defines incoherent ergotropy $W_{\rm inc}(\rho)$ of $\rho$.
\end{definition}
The incoherent ergotropy $W_{\rm inc}$ can be interpreted as the maximal work that can be extracted from $\rho$ by permutating its energetic populations. Therefore, it can be equivalently  characterized as the ergotropy of the decohered state $\rho_{\rm diag} :  =  \sum_k   \,  \<  \epsilon_k| \rho  |\epsilon_k\>  \,  |\epsilon_k\>\<\epsilon_k|$, obtained by setting to zero all the off-diagonal elements of $\rho$ in the energy basis ~\cite{Francica2020}:
\begin{align}
W_{\rm inc}  (\rho) =  W  (\rho_{\rm diag}) \,.       
\end{align}
For qubits, it takes the simple expression 
\begin{align}\label{eq:IncohErg}
W_{\rm inc}  (\rho) =  \max\Big\{0,   \Delta E  \,   \big(  \rho_{11}  -  \rho_{00}\big)   \Big\} \,,
\end{align}
with $\Delta E  := \epsilon_1  - \epsilon_0$ and $\rho_{kk}  :=\< \epsilon_k| \rho |\epsilon_k\>$ for $k\in  \{0,1\}$, showing that a quantum state has incoherent ergotropy if and only if it exhibits  population inversion in the energy basis.     

The rest of ergotropy, i.e., the difference between $W(\rho)$ and $W_{\rm inc}(\rho)$ is associated with the work that can be extracted by consuming the coherence stored in $\rho$. For this reason, it is denoted as \textit{coherent ergotropy}~\cite{Francica2020}, as highlights the following Definition.
\begin{definition}[Coherent ergotropy]
    Given a work storage $S$ in state $\rho$, the difference between its ergotropy $W(\rho)$ and incoherent ergotropy $W_{\rm inc}(\rho)$ defines the coherent ergotropy of $\rho$:
    \begin{equation}
        W_{\rm coh}(\rho) :  = W(\rho) - W_{\rm inc}(\rho).
    \end{equation}
\end{definition}
For qubits, the coherent ergotropy has the simple expression~\cite{Francica2020} 
\begin{equation}\label{eq:CohErg}
        W_{\rm coh}(\rho) = \frac{1}{2}\left(\eta-\sqrt{\eta^2 - 4 |\rho_{01}|^2}\right)\, ,
    \end{equation}
    where $\rho_{01}$ is the off-diagonal element $ \rho_{01}: =\<  \epsilon_0 |\rho|\epsilon_1\>$,    
     $\eta:=\sqrt{2P(\rho)-1}$, and $P(\rho)$ is the purity  $P(\rho):=\operatorname{Tr}(\rho^2)$. Note that the coherent ergotropy is non-zero if and only if the state has non-zero coherence between the two energy levels. 
  
  In this paper, we focus on a qubit system $S$ characterized by Hamiltonian $H_S = \epsilon_0 |\epsilon_0\rangle \langle \epsilon_0 | + \epsilon_1 |\epsilon_1\rangle \langle \epsilon_1 |$ with $\epsilon_0 \leq \epsilon_1$. Without loss of generality, the corresponding energies can be rescaled as $\epsilon_0 = 0$ and $\epsilon_1 = 1$: therefore, in Sections~\ref{sec:switch} and~\ref{sec:corr}, we consider a system $S$ with a Hamiltonian
  \begin{equation}\label{refHam}
      H_S = |1\rangle\langle 1|.
  \end{equation}

\subsubsection{Ergotropy and correlations with ancilla.}

In the previous sections, we have identified how a given state $\rho$ of a quantum system can be used as fuel to perform work by manipulating its Hamiltonian. So far, we treated the system as isolated. However, taking into account the presence of an additional ancilla $C$ correlated with the system $S$ may cause an enhancement of the extractable work from $S$ itself. For instance, suppose that the work medium and ancilla are in a joint state $\rho_{SC}$ with $\operatorname{Tr}_C[\rho_{SC}] = \rho$. Although they are not jointly accessible, the outcome of a measurement performed on $C$ can be classically communicated to the party holding $S$. Mathematically, it can be described by the corresponding positive operator-valued measure (POVM) $\pmb{\Theta} = \{ \Theta_i \}$ preparing the work medium $S$ in a state
\begin{equation}
 \rho_{S|i}  :  =\frac{1}{p_i} \operatorname{Tr}_C[   (\mathds{1}_S \otimes \Theta_i)  \rho_{SC} ],
 \end{equation}
carrying ergotropy $W(\rho_{S|i})$ according to~(\ref{eq:ergotropy}), with probability $p_i = \operatorname{Tr}[   (\mathds{1}_S \otimes \Theta_i)  \rho_{SC} ]$. If the outcome of the measurement is not known, the work medium remains in the state $\rho = \sum_i p_i \rho_{S|i} \equiv \operatorname{Tr}_C[\rho_{SC}]$ carrying ergotropy 
\begin{equation}\label{locErg}
W_{\rm loc}(\rho_{SC}) = W(\operatorname{Tr}_C[\rho_{SC}]),    
\end{equation}
which we denote as local. However, if the outcome of $\pmb{\Theta}$ is communicated, one can extract from the state of $S$ the ergotropy $\sum_i p_i W(\rho_{S|i})$ averaged over the possible outcomes. In many realistic scenarios, including ours, the amount of classical communication between the ancilla $C$ and the system of interest $S$ is limited. Therefore, we restrict $C$ to communicate only one bit of information. This restricts the measurements $\pmb{\Theta}$ of the control system to be two-outcome ones. In turn, any two-outcome POVM can be simulated by a mixture of projective measurements~\cite{Masanes2005, Masanes2006}. Taking into account convexity of ergotropy $\sum_i p_i W(\rho_{S|i})$~\cite{Bernards2019}, it guarantees that the maximal work can be extracted from $S$ with assistance of projective measurements on $C$. Therefore, in what follows, we restrict our analysis to projective measurements $\pmb{\Theta}$, and optimization of resulting ergotropy over all possible projective measurements $\pmb{\Theta}$ on $C$ determines the notion of {\em daemonic ergotropy}~\cite{Francica2017}.
    \begin{definition}[Daemonic ergotropy]\label{def:DaemonicErg}
        Daemonic ergotropy of the work storage $S$ with respect to an ancilla $C$ is the maximal average ergotropy that can be stored in $S$ after a projective measurement of $C$ with post-selection:
        \begin{equation}
        \label{DaemonicErg}
            W_{\rm dae}(\rho_{SC}) := \max_{ \pmb{\Theta} } \Bigl(\sum_i p_i W(\rho_{S|i})\Bigr) \, .
        \end{equation}    
    \end{definition}
Daemonic ergotropy catches the influence of correlations between $S$ and $C$ on work extraction since they imply a gap between $W_{\rm dae}$ and $W_{\rm loc}$~\cite{Francica2017}.

\subsection{Thermal states and operations}\label{sec:thermalStates}
  
An important example of passive states is provided by the thermal states (a.k.a. Gibbs states), of the form  
 \begin{equation}\label{eq2}
\tau_{S, \beta} = \frac{e^{-\beta H_S}}{Z_{S,\beta}} \,,
\end{equation}
where $\beta  :=  (\kappa_B T)^{-1}$ is the inverse temperature ($\kappa_B$ being Boltzmann's constant), and $Z_{S, \beta} = \text{Tr}[e^{-\beta H_S}] = \sum_{k}e^{-\beta \epsilon_{k}}$ is the canonical partition function. Indeed, they are diagonal in the Hamiltonian eigenbasis, and their eigenvalues  
\begin{align}\label{gibbseigen} 
p_k   =  \frac{e^{- \beta    \epsilon_{k}}}{Z_{S, \beta}}
\end{align}
are monotonically decreasing with respect to  the energy of the corresponding levels.  In general, however, not all passive states are Gibbs: the Gibbs states can be characterized as the {\em completely passive} states, {\em i.e.} the states  $\rho$ such that $\rho^{\otimes n}$ is passive for every possible integer $n\in  \mathbb{N}$~\cite{Lenard1978,Pusz1978}. In this paper, we will restrict our attention to the qubit case, where every passive state can be regarded as a Gibbs state at a suitable temperature $T$.  Hence, the difference between passive and completely passive states will not play a role in our results.

Complete passivity of thermal states suggests that, in the presence of suitable environment $E$, they can be created at no work cost and, therefore, can be considered from resource-theoretic perspective as free states. In addition to that, we can consider a set of operations that can be implemented on a system $S$ with no cost. For example, given $S$ interacting with environment $E$ in a thermal state $\tau_{E, \beta} = \frac{e^{-\beta H_E}}{Z_{E,\beta}}$ (with $Z_{E,\beta} = \operatorname{Tr}[e^{-\beta H_E}]$) at temperature $T = (\kappa_B \beta)^{-1}$, conservation of total energy sets certain restrictions on their joint unitary evolution $U_{SE}$, namely, $[U_{SE}, H_S \otimes \mathds{1}_E + \mathds{1}_S \otimes H_E] = 0$, where $H_E$ is the Hamiltonian of $E$, and $\mathds{1}_S$ and $\mathds{1}_E$ denote the identity operators on the Hilbert spaces of $S$ and $E$, respectively. If the state of $S$ is also thermal at temperature $T$ and non correlated with $E$, such energy-preserving unitaries do not affect both states, $U_{SE} (\tau_{S, \beta} \otimes \tau_{E, \beta}) U_{SE}^\dagger = \tau_{S, \beta} \otimes \tau_{E, \beta}$. Therefore, they can be seen as ones generating a set of free operations, namely, \textit{thermal operations} $\mathcal{T}$~\cite{Janzing2000, Brandao2013, Horodecki2013},
\begin{equation}\label{eq:ThermalOp}
\mathcal{T}[\rho]=\text{Tr}_{E}[U_{SE}(\rho\otimes \tau_{E, \beta})U_{SE}^{\dagger}],
\end{equation}
where $\operatorname{Tr}_E$ is the partial trace over the environment's Hilbert space. Thermal operations preserve the thermal state of the system,
\begin{equation}\label{eq:GibbsPreserv}
\mathcal{T}[\tau_{S,\beta}] = \tau_{S,\beta},
\end{equation}
and, therefore, cannot generate resource on its own without the corresponding supply.

A radical example of thermal operations are the {\em thermalizing channels} which replace the initial state of the system with the thermal state,
\begin{equation}\label{eq:ThermalChannel}
\map T_{\beta}[\rho]=\tau_{S,\beta}  \qquad \forall \rho  \, ,
\end{equation}
hence, trivially satisfying the condition (\ref{eq:GibbsPreserv}) of preservation of $\tau_{S,\beta}$. Technically, they can be considered as a generalization of the completely depolarizing channel that outputs a maximally mixed state regardless of the input \cite{Felce2020}: in fact, the latter is a thermalizing channel at infinite temperature,
\begin{equation}
\map T_{\beta = 0}[\rho] = \frac{\mathds{1}_S}{d}  \qquad \forall \rho  \, ,
\end{equation}
where $d$ is the dimension of the Hilbert space of $S$. From the thermodynamic point of view, thermalizing channels (\ref{eq:ThermalChannel}) can be seen as modelling full thermalization of $S$, which is coupled with the environment $E$ at fixed temperature $T = (\kappa_B \beta)^{-1}$, via relaxing it to the corresponding thermal state \cite{deOliveira2020, ChapeauBlondeau2022}. For the considered qubit system $S$ with the Hamiltonian~(\ref{refHam}), its interaction with $E$ can be interpreted as action of a thermal noise channel $\map G [\rho]  =  \sum_{i,j}   G_{ij}   \,  \rho  G_{ij}^\dag$, with the Kraus operators
\begin{align}\label{eq:ThermKrausG}
G_{00} &= \sqrt{p}\Bigl(\ket{0}\bra{0} + e^{-\frac{\lambda}{2}t} \ket{1}\bra{1} \Bigr), \nonumber \\
G_{01} &= \sqrt{p(1 - e^{-\lambda t})}\ket{0}\bra{1}, \nonumber \\
G_{10} &= \sqrt{(1-p)(1 - e^{-\lambda t})}\ket{1}\bra{0}, \nonumber \\
G_{11} &= \sqrt{1-p}\Bigl(e^{-\frac{\lambda}{2}t} \ket{0}\bra{0} + \ket{1}\bra{1} \Bigr),
\end{align}
where $p = (1+e^{-\beta})^{-1} \in [1/2,1]$ is taken with respect to Hamiltonian~(\ref{refHam}), $t$ is the interaction time, and $\lambda$ is the damping constant defining characteristic relaxation time \cite{Jevtic2015, Tham2016, Mancino2017, deOliveira2020, ChapeauBlondeau2022}. Given $\rho = \begin{pmatrix} \rho_{00} & \rho_{01} \\ \rho_{01}^* & \rho_{11} \end{pmatrix}$, the system $S$ is left in the state
\begin{equation}
    \mathcal{G}[\rho] = \begin{pmatrix} e^{-\lambda t} \rho_{00} + (1 - e^{-\lambda t}) p & e^{-\lambda t} \rho_{01} \\ e^{-\lambda t} \rho_{01}^* & e^{-\lambda t} \rho_{11} + (1-e^{-\lambda t})(1-p) \end{pmatrix}
\end{equation}
after the action of the thermal noise channel. In turn, for long times $t \gg \lambda^{-1}$, the thermal noise channel reduces to thermalizing channel $\map T_\beta  [\rho]  =  \sum_{i,j}   T_{\beta, ij}   \,  \rho  T_{\beta,  ij}^\dag$, with the Kraus operators
\begin{align}\label{eq:ThermKraus}
T_{\beta, 00} &= \sqrt{p}\ket{0}\bra{0}, \nonumber \\
T_{\beta, 01} &= \sqrt{p}\ket{0}\bra{1}, \nonumber \\
T_{\beta, 10} &= \sqrt{1-p}\ket{1}\bra{0}, \nonumber \\
T_{\beta, 11} &= \sqrt{1-p}\ket{1}\bra{1},
\end{align}
achieving the desired thermal state $\tau_{S, \beta} = \frac{1}{1 + e^{-\beta}} \begin{pmatrix} 1 & 0 \\ 0 & e^{-\beta}\end{pmatrix}$ as required by (\ref{eq:ThermalChannel}).
\par

\section{Thermalization in a coherently controlled order}\label{sec:switch}

\subsection{Environmental influence and thermalization}\label{subsec:cassep}

\begin{figure}[t]
    \centering
    \includegraphics[width = \linewidth]{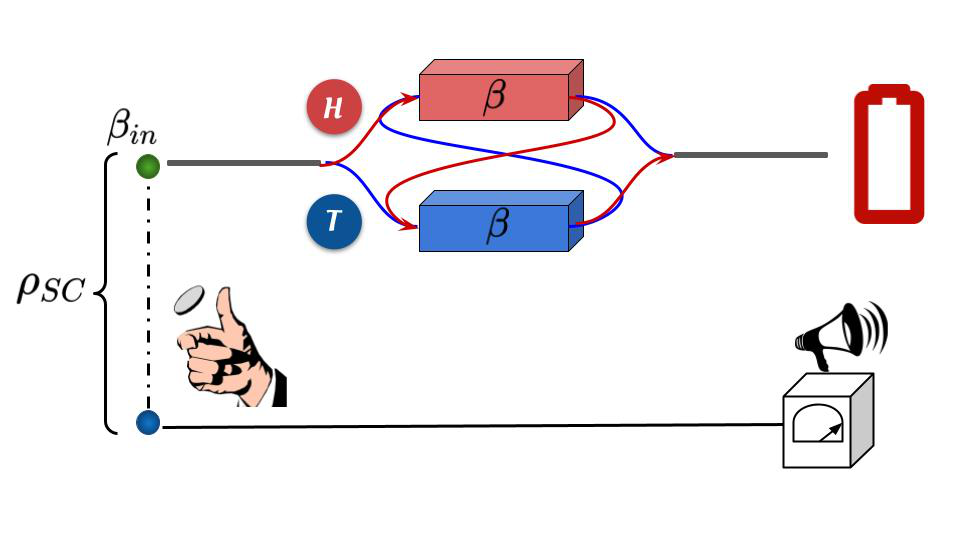}
    \caption{Schematic representation of environmental influence modelled by two identical channels. Any causally separable order (well-defined or probabilistic) of them outputs a thermal state of $S$ storing no work and destroys initial correlations with ancilla $C$ if any.}
    \label{fig:ClassConfig}
\end{figure}

In the ergotropic treatment of work, an active state of the work storage $S$ is taken as granted without relying on its origin. However, in realistic scenarios, $S$ can be influenced by a (thermal) environment that causes degradation of the carried thermodynamic resource. For example, $S$ can be transmitted between the laboratories via noisy channels or interact with external systems such as thermal baths or measurement apparatus, not necessarily in a well-defined order~\cite{Guha2020, Felce2020, Simonov2022, Nie2022, Dieguez2023}. The influence of thermal environment on $S$ can be modelled by thermal operations~(\ref{eq:ThermalOp}): Here, for the sake of simplicity, we focus on the set of thermalizing channels~(\ref{eq:ThermalChannel}), assuming hence complete thermalization with the environment. Therefore, before ergotropic work is extracted from the storage $S$, it is affected by interaction with the environment, which we model by a sequence of $N$ thermalizing channels $\mathcal{T}_{\beta_1}, ..., \mathcal{T}_{\beta_N}$. For any possible order in which the channels act, $S$ is found in a thermal state at temperature associated with the last thermalizing channel,
\begin{equation}\label{eq:WellDefOrder}
    \rho = (\mathcal{T}_{\beta_{i_1}} \circ ... \circ \mathcal{T}_{\beta_{i_N}})[\rho_{\rm in}] = \tau_{\beta_{i_N}},
\end{equation}
which obviously carries zero ergotropy. In turn, the order of channels can depend on an external degree of freedom given by the ancilla $C$, which is not necessarily accessible for the holder of $S$. For example, let us assume that $C$ chooses between $N$ cyclic orders of $\mathcal{T}_{\beta_1}, ..., \mathcal{T}_{\beta_N}$ with probabilities $(p_i)_i$. In this case, the joint state of both $S$ and $C$ is transformed as
\begin{eqnarray}\label{eq:PinMap}
    \nonumber \rho_{SC} &\rightarrow& \Big[\Bigr( \sum_i p_i \mathcal{P}_i(\mathcal{T}_{\beta_1}, ..., \mathcal{T}_{\beta_N}) \Bigr) \otimes \map I_C \Big] \, [\rho_{SC}] \\
    &=:& \sigma^{\rm sep}_{N, (\beta_i)_i, \rho_{\rm in}},
 \end{eqnarray}
with $p_i \in [0,1]$ and $\sum_i p_i = 1$, and $\mathcal{I}_C$ being an identity map on the space of $C$, and where index $N$ corresponds to the number of thermalizing channels, while index ``sep'' highlights that these are applied in a causally separable order. According to~(\ref{eq:WellDefOrder}), any causally separable order of $\mathcal{T}_{\beta_1}, ..., \mathcal{T}_{\beta_N}$ produces a fixed thermal output state irrespective of the initial state of $S$ (i.e., realizes a pin map). Therefore, even if the work storage $S$ initially shares correlations with $C$, they are destroyed by action of the pin map~(\ref{eq:PinMap}). Hence, if $C$ is not accessible, $S$ remains in a probabilistic mixture of thermal states at temperatures of each channel,
\begin{equation}\label{eq:CausSepOrder}
    \rho = \sum_{i=1}^N p_i \tau_{\beta_i}.
\end{equation}
It establishes again a completely passive state carrying zero ergotropy, meaning that local ergotropy of $S$ with respect to $C$ is zero according to~(\ref{locErg}),
\begin{equation}
    W_{\rm loc}(\sigma^{\rm sep}_{N, (\beta_i)_i, \rho_{\rm in}}) = 0.
\end{equation}
On the other hand, since $ \sigma^{\rm sep}_{N, (\beta_i)_i, \rho_{\rm in}}$ does not carry any correlations between $S$ and $C$, a measurement occurring on $C$ with its outcome known to the holder of $S$ prepares the latter again in~(\ref{eq:CausSepOrder}). Hence, daemonic ergotropy of $S$ with respect to $C$ is also zero according to~(\ref{DaemonicErg}),
\begin{equation}
    W_{\rm dae}(\sigma^{\rm sep}_{N, (\beta_i)_i, \rho_{\rm in}}) = 0,
\end{equation}
as shown on Fig.~\ref{fig:ClassConfig}.

Therefore, thermalizing channels acting in a causally separable order (i.e., in any well-defined order or probabilistic mixture of them) nullify the work initially stored in $S$ and output a passive state. In what follows, we question whether the state of $S$ remains passive if the assumption of the causally separable order of thermal environment's influence on $S$ is relaxed. Before to proceed with the results, we provide the reader with the framework necessary to describe indefinite causal order of thermalizing channels.

\subsection{Quantum SWITCH: Indefinite causal order of channels}

Quantum mechanics allows for scenarios that go beyond causally separable combinations of multiple thermalization processes considered above, making their causal order subject to quantum uncertainty. A particular scenario of this kind is provided by the quantum SWITCH~\cite{chiribella2009beyond, Chiribella2013}, a higher-order operation acting on quantum channels, which realizes coherent control of their order making it incompatible with any possible causally separable combination of them. Assuming the ancilla $C$ to be a quantum $2$-dimensional system (i.e., a qubit), the simplest version of the quantum SWITCH takes in input two arbitrary quantum channels $\map E$ and $\map F$, acting on $S$, and produces in output a bipartite channel $\map S  (\map E, \map F)$, acting on $S$ and $C$. Notice that, in this case, the ancilla $C$ acts as the control of the quantum SWITCH. Thus, from a terminological perspective, in what follows, we refer to $C$ as ancilla or control system interchangeably.
\begin{definition}[Quantum SWITCH of $2$ channels]
    The quantum SWITCH controlling the order of two quantum channels $\map E$ and $\map F$ is a supermap $\mathcal{S}$ that assigns to them a new bipartite quantum channel
    \begin{align}\label{eq:switchchannel}
\big[ \map S  (\map E,\map F) \big] \, (\rho_{SC}) =  \sum_{i,j}  \,  S_{ij}  \,  \rho_{SC}  \,  S_{ij}^\dag  \, , 
\end{align}
where $S_{ij}$ is the Kraus operator defined by 
\begin{align}\label{switchkraus}
S_{ij}  :  =   E_i  F_j \otimes |0\>\<0|_C   + F_j E_i \otimes |1\>\<1|_{C} \, ,  
\end{align}
 $\{|0\>_C,  |1\>_C  \}$ being an orthonormal basis for the control qubit, and $\{E_i\}$ and $\{F_j\}$ being two Kraus representations of channels $\map E$ and $\map F$, respectively.
\end{definition}
Crucially, the quantum channel $\map S (\map E, \map F)$ defined in Eq.~(\ref{eq:switchchannel}) is independent of the choice of Kraus representations used for channels $\map E$ and $\map F$. In practice, one can choose whatever Kraus representation is more convenient or more insightful in the problem at hand. In turn, a measurement performed on $C$ prepares $S$ in a state, which is not necessarily compatible with any causally separable combination of $\map E$ and $\map F$.
 
If the system $S$ and control qubit $C$ initially share no correlations, i.e., $\rho_{SC} = \rho_{\rm in} \otimes \omega$ with $\rho_{\rm in}$ and $\omega$ being states of $S$ and $C$, respectively, the channel $\map S  (\map E,  \map F)$ outputs
\begin{align}\label{uncorrswitch}
    \nonumber \big[ \map S (\map{E}, \map{F}) \big]  (\rho_{\rm in} \otimes \omega )  = \;  & \frac{1}{4} \sum_{ij} \Bigl( \{E_i, F_j\} \rho_{\rm in} \{E_i, F_j\}^\dagger \otimes \omega  \\
    \nonumber &+  \{E_i, F_j\} \rho_{\rm in} [E_i, F_j]^\dagger \otimes \omega Z \\
    \nonumber &+ [E_i, F_j] \rho_{\rm in} \{E_i, F_j\}^\dagger \otimes Z\omega \\
    &+ [E_i, F_j] \rho_{\rm in} [E_i, F_j]^\dagger \otimes Z\omega Z \Bigr),
\end{align}
where  $[E_i,F_j]:=  E_iF_j  -  F_j  E_i$ denotes the commutator,  $\{E_i,F_j\}:  =   E_i  F_j  +  F_j E_i$ denotes the anti-commutator, and $Z:  = |0\>\<0| -|1\>\<1|$. Performance of a measurement on $C$ prepares $S$ in one of the states not necessarily compatible with any causally separable combination of $\map E$ and $\map F$. In turn, their average ergotropy optimized over all possible $\omega$ and measurements of $C$ defines the daemonic ergotropy that can be associated with enhancement of extractable work brought by the quantum SWITCH.

The considered construction of the quantum SWITCH can be straightforwardly generalized to a higher-order operation acting on $N \geq 2$ quantum channels~\cite{Procopio2019, Procopio2020, Chiribella2021L}. Indeed, it takes in input $N$ arbitrary quantum channels $\map E_1, \cdots, \map E_N$, which act on $S$, and produces in output a bipartite channel $\map S  (\map E_1, \cdots, \map E_N)$, acting on both the target system $S$ and control system $C$, which is now represented by a suitable quantum $N$-level system.

\begin{definition}[Quantum SWITCH of $N \geq 2$ channels]
    The quantum SWITCH controlling cyclic orders of $N$ quantum channels $\map E_1, \cdots, \map E_N$ is a supermap $\mathcal{S}$ that assigns to them a new bipartite quantum channel
\begin{eqnarray}\label{eq:corrNSwitchAction}
  \big[\mathcal{S}(\map E_{1}, \cdots, \map E_{N})\big] (\rho_{\rm{SC}}) = \sum_{i_{1},\cdots,i_{N}}S_{i_{1}\cdots i_{N}}(\rho_{\rm{SC}}) S_{i_{1}\cdots i_{N}}^{\dagger},
\end{eqnarray}
where $S_{i_1 {\cdots} i_N}$ are the Kraus operators given by
\begin{equation}\label{eq:NSwitchKraus}
    S_{i_{1}{\cdots} i_{N}} := \sum_{j=0}^{N-1}\mathcal{P}_{j}\Bigl(E_{i_{1}}^{(1)}{\cdots} E_{i_{N}}^{(N)}\Bigr)_{\rm{S}}\otimes\ket{j}\bra{j}_{\rm{C}},
\end{equation}
with $\{E_{l}^{(k)}\}$ being a Kraus representation of a channel $\map E_{k}$, while $\mathcal{P}_{j}$ represents a $j$-cyclic permutation of Kraus operators associated with different channels.
\end{definition}
In analogy with~(\ref{switchkraus}), in the quantum $N$-SWITCH, the computational basis states of $C$ trigger different \textit{cyclic} orders of $\map E_1, \ldots, \map E_N$.

\subsection{Work extraction under thermalization in an indefinite causal order}\label{sec:ThermoSwitch}

\begin{figure}[t]
    \centering
    \includegraphics[width = \linewidth]{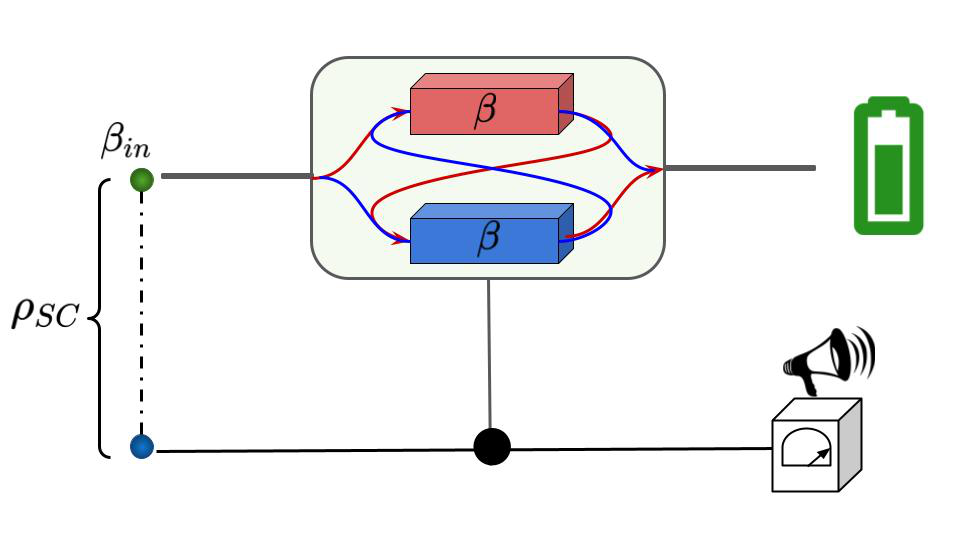}
    \caption{Schematic representation of environmental influence via thermalizing channels $\mathcal{T}_\beta$ in indefinite causal order. The work medium $S$ and control system $C$ are initially in a joint state $\rho_{SC}$, and the former is locally at temperature $\beta_{\rm in}$, i.e., $\operatorname{Tr}_C[\rho_{SC}] = \tau_{\beta_{\rm in}}$. $S$ undergoes action of thermalizing channels (here, the case $N=2$ is shown) of inverse temperature $\beta$ whose application order is controlled by the quantum SWITCH. Thereafter, $C$ undergoes a measurement, and the ergotropic work is extracted from the resulting state of $S$.}
    \label{fig:Setup}
\end{figure}

Now, we turn back to the scenario of work extraction from $S$, which is previously influenced by the environment (in what follows, we choose units such that $\hbar = \kappa_B = 1$). It can be summarized in the following steps:
\begin{itemize}
\item Work storage $S$ and the ancilla $C$ are initially described by  the joint state $\rho_{SC}$, and the marginal state of the former is resourceless, i.e., a thermal state $\tau_{\beta_{\rm in}}$ at temperature $T_{\rm in} = \beta_{\rm in}^{-1}$,
\begin{equation}\label{eq:LocalThermality}
    \operatorname{Tr}_C (\rho_{SC}) = \tau_{\beta_{\rm in}}.
\end{equation}
\item Work storage $S$ is influenced by the environment via a sequence of thermalizing channels \eqref{eq:ThermalChannel} whose causal order is determined by $C$. Specifically, $C$ functions as the control system of the quantum SWITCH described in \eqref{eq:corrNSwitchAction}, providing coherent control over the order of thermalization processes, each involving interaction with the thermal environment and resulting in $S$ reaching a thermal state.
\item Finally, daemonic ergotropy \eqref{DaemonicErg} is extracted from $S$ with the assistance of the ancilla $C$.
\end{itemize} 
Here we assume that the same temperature $T = \beta^{-1}$ is associated with all the channels, i.e., $S$ undergoes action of $N \geq 2$ identical channels, each described by the Kraus operators \eqref{eq:ThermKraus} and leaving it in the thermal state $\tau_\beta$. Therefore, our figure of merit is the maximal work that can be extracted from $S$ after the action of the thermalizing channels $\mathcal{T}_\beta$ whose order is controlled by the quantum SWITCH, as shown on Fig.~\ref{fig:Setup}. It can be formalized by the following quantity.

\begin{definition}[SWITCH-ergotropy]
    The SWITCH-ergotropy is the daemonic ergotropy of the work storage $S$ with respect to the control system $C$ after the action of the quantum SWITCH controlling the order of $N \geq 2$ thermalizing channels $\mathcal{T}_\beta$:
    \begin{equation}\label{def:SwitchErg}
        W_{\mathcal{S}}^N(\rho_{SC}) = W_{\mathrm{dae}}\Bigl( \big[\mathcal{S}(\underbrace{\mathcal{T}_{\beta}, ..., \mathcal{T}_{\beta}}_{N\mathrm{ channels}})\big] (\rho_{SC}) \Bigr).
    \end{equation}
\end{definition}
In what follows, we analyse the SWITCH-ergotropy~(\ref{def:SwitchErg}) for several configurations of initial joint state $\rho_{SC}$ of the work storage and control system starting with uncorrelated scenario and proceede with the case of $S$ and $C$ initially sharing classical and quantum correlations, respectively.

\subsubsection{Indefinite causal order of 2 thermalizing channels}

We start by reviewing the simplest case of $S$ undergoing action of $N = 2$ identical channels $\mathcal{T}_\beta$~\cite{Simonov2022}, whose order is controlled by the quantum SWITCH, as shown in Fig.~\ref{fig:Setup}. It is assumed that the work storage is initially uncorrelated with $C$ and prepared in a thermal state $\rho_{\rm in} \equiv \tau_{\beta_{\rm in}}$ at temperature $T_{\rm in} = \beta_{\rm in}^{-1}$,
\begin{equation}\label{eq:InitStateUncorr}
    \rho_{SC} = \tau_{\beta_{\rm in}} \otimes \omega.
\end{equation}
Therefore, it carries zero ergotropy as well as zero daemonic ergotropy before the action of thermalizing channels. Obviously, if $\mathcal{T}_\beta$ act in a causally separable order, the same holds for the final state in accordance with~(\ref{eq:CausSepOrder}). On the other hand, the SWITCH-ergotropy, generally speaking, is not necessarily zero and depends on the initial state $\omega$ of the control qubit. Therefore, we focus on the maximal SWITCH-ergotropy with respect to all quantum states $\omega$ providing the following Definition.
\begin{definition}[Maximal SWITCH-ergotropy without a priori correlations]
    The maximal SWITCH-ergotropy without a priori correlations is the SWITCH-ergotropy with respect to $N$ thermalizing channels $\mathcal{T}_\beta$ optimized over all uncorrelated states~(\ref{eq:InitStateUncorr}):
    \begin{equation}
        \mathcal{W}_{\mathrm{UC}}^{N} := \max_\omega W_{\mathcal{S}}^N (\tau_{\beta_{\rm in}} \otimes \omega).
    \end{equation}
\end{definition}
First, we report the result for the maximal SWITCH-ergotropy without a priori correlations with respect to $N=2$ thermalizing channels.
\begin{theorem}\label{theor:ErgUC2}
    The maximal SWITCH-ergotropy without a priori correlations with respect to two thermalizing channels $\mathcal{T}_\beta$ is provided by the optimal state:
    \begin{equation}
        \rho_{SC}^{\mathrm{opt}} = \tau_{\beta_{\rm in}} \otimes |+\rangle\langle +|
    \end{equation}
    and given by:
    \begin{equation}\label{eq:2Erg}
        \mathcal{W}_{\mathrm{UC}}^{N=2} = \frac{\operatorname{max}\{ 0, e^{-2\beta} - e^{-\beta_{\rm in}}\}}{2Z_{S,\beta}^2Z_{S,\beta_{\rm in}}},
    \end{equation}
    which is purely incoherent.
    \begin{proof}
        See Appendix~\ref{appUC2}.
    \end{proof}
\end{theorem}

The maximal SWITCH-ergotropy~(\ref{eq:2Erg}) of $S$ represents the maximal work that can be extracted from it when $\mathcal{T}_\beta$ act in an indefinite causal order. Indeed, it is not necessarily zero and has a genuinely incoherent origin, i.e., can be extracted by permutating the energetic populations. Importantly, the condition for a non-zero~(\ref{eq:2Erg}) establishes a \textit{temperature bound} at which thermalization in a quantum-controlled order activates the initially passive state of the working medium $S$, which we provide in the following Corollary.
\begin{corollary}
    The maximal SWITCH-ergotropy~(\ref{eq:2Erg}) without a priori correlations is non-zero under condition
    \begin{equation}\label{eq:TempLimit}
    \beta_{\rm in} > 2 \beta,
    \end{equation}
    which we refer to as the temperature bound\footnote{A similar temperature bound $\beta_{\rm in} > \beta_1 + \beta_2$ can be derived for quantum-controlled thermalizing channels at different inverse temperatures $\beta_{1,2}$. Nevertheless, the optimal value of ergotropy is obtained for the case of channels at equal temperatures, $\beta_1 = \beta_2 = \beta$, therefore, in what follows, we stick to this setting.}.
    \begin{proof}
        The condition $\mathcal{W}_{\mathrm{UC}}^{N=2} > 0$ leads to a condition $e^{-2\beta} - e^{-\beta_{\rm in}} > 0$, from which follows the temperature bound~(\ref{eq:TempLimit}).
    \end{proof}
\end{corollary}

\subsubsection{Indefinite causal order of N thermalizing channels}

 A natural question arises whether the temperature bound~(\ref{eq:TempLimit}) can be shifted or overcome. At first, one can ask whether it can be done by increasing the number of the thermalizing channels $\mathcal{T}_\beta$ acting on $S$. Indeed, in analogy with the above quantum SWITCH scenario, let us assume that $N \geq 2$ channels at temperature $T$ act on $S$ being prepared initially in a thermal state $\rho = \tau_{\beta_{\rm in}}$ at temperature $T_{\rm in}$.
\begin{theorem}\label{theor:ErgUCN}
    The maximal SWITCH-ergotropy without a priori correlations with respect to $N\geq 2$ thermalizing channels $\mathcal{T}_\beta$ is provided by the optimal state:
    \begin{equation}
        \rho_{SC}^{\mathrm{opt}} = \tau_{\beta_{\rm in}} \otimes |\gamma_+\rangle\langle \gamma_+|,
    \end{equation}
    where $\ket{\gamma_+} = \frac{1}{\sqrt{N}} \sum_{i=0}^{N-1} \ket{i}$, according to~(\ref{eq:corrNSwitchAction}), and given by:
     \begin{equation}\label{eq:NErg}
        \mathcal{W}_{\mathrm{UC}}^{N} = \frac{N-1}{NZ_{S,\beta}^2Z_{S,\beta_{\rm in}}} \operatorname{max}\{ 0, e^{-2\beta} - e^{-\beta_{\rm in}} \},
    \end{equation}
    which is purely incoherent.
    \begin{proof}
        See Appendix~\ref{appUCN}.
    \end{proof}
\end{theorem}
This means that for the initial temperatures $\beta_{\rm in}$ satisfying the bound~(\ref{eq:TempLimit}), the work stored in $S$ after action of $N$ thermalizing channels in a causally non-separable order can be improved by increasing their number,
\begin{equation}
    \mathcal{W}_{\mathrm{UC}}^{N} = 2\Bigl(1 - \frac{1}{N} \Bigr) \mathcal{W}_{\mathrm{UC}}^{N=2},
\end{equation}
up to doubling the work which can be extracted in the usual scenario of quantum SWITCH controlling the order of two channels. However, the temperature bound~(\ref{eq:TempLimit}) remains unchanged and cannot be beaten in this manner.

\section{Controlled thermalization assisted by prior correlations}\label{sec:corr}

The results obtained above suggest that, although quantum SWITCH allows one to activate a thermal state by quantum-controlled thermalization, it reveals a constraint given by the condition~(\ref{eq:TempLimit}) on its temperature which does not depend on the number of controlled thermalizing channels $N$. Therefore, it should be asked whether it can be overcome differently, for example, by allowing initial correlations between the work storage $S$ and the control system $C$. 
To this aim, for the sake of simplicity, we consider again the quantum SWITCH of $N=2$ channels and weaken the condition~(\ref{eq:InitStateUncorr}), hence, demanding $S$ to be in a thermal state only locally with respect to \eqref{eq:LocalThermality}. 

Note that, although the joint state $\rho_{SC}$ is locally passive, it can carry non-zero daemonic ergotropy. However, crucially, as we highlighted in Section~\ref{subsec:cassep}, sharing a priori correlations (classical as well as quantum) between the work medium and control does not help one to extract work from the former if the maps act consecutively or in mixture of their causal orders (for example, when the control system is discarded). 

Interestingly, for an initially uncorrelated joint state~(\ref{eq:InitStateUncorr}), the SWITCH-ergotropy (if any) $W_\mathcal{S}(\rho_{SC})$ can be interpreted as ergotropic gain or generation. On the other hand, if $\rho_{SC}$ initially shares some correlations under the condition~(\ref{eq:LocalThermality}) of local thermality, the SWITCH-ergotropy can be interpreted in two distinct ways. If it exceeds the initial daemonic ergotropy, i.e., $ W_\mathcal{S}(\rho_{SC}) > W_\mathcal{\mathrm{dae}}(\rho_{SC})$, it corresponds to ergotropic generation as in the case of no a priori correlations. Otherwise, it can be interpreted as preservation of ergotropy by the quantum SWITCH.

\subsection{Classical correlations between work medium and control}

We start the analysis with the case of work storage $S$ and control system $C$ sharing prior classical correlations. Importantly, the initial correlations are assumed to be genuinely classical, i.e., the joint state of $S$ and $C$ carries no quantum discord~\cite{discord1, discord2}. Hence, we can write the system-control joint state as
\begin{equation}\label{eq:ClassicalExpansion}
\rho_{SC} = \sum_{ij} p_{ij} \Pi_i \otimes \Theta_j,
\end{equation}
where, both $\{\Pi_i\}_i$ and $\{\Theta_j\}_j$ are the rank-1 orthogonal projectors on their respective subsystems, and $p_{ij}$ is the probability corresponding to the state $\Pi_i \otimes \Theta_j$,

The condition~(\ref{eq:LocalThermality}) of local thermality puts constraints on the possible expansions~(\ref{eq:ClassicalExpansion}) of the joint state $\rho_{SC}$, and we introduce the following Definition.
\begin{definition}
    The maximal SWITCH-ergotropy under initial classical correlations is the SWITCH-ergotropy with respect to two thermalizing processes $\mathcal{T}_\beta$ optimized over all initial classically correlated states~(\ref{eq:ClassicalExpansion}):
    \begin{equation}\label{eq:Erg2CCgen}
        \mathcal{W}_{\rm CC} := \max_{\substack{\{p_{ij}, \Pi_i, \Theta_i\} \\ \sum_{ij} p_{ij} \Pi_i = \tau_{\beta_{\mathrm{in}}} }} W_{\mathcal{S}}^{N=2}\Big(\sum_{ij} p_{ij} \Pi_i \otimes \Theta_j\Big),
    \end{equation}
    under the condition~(\ref{eq:LocalThermality}) of local thermality.
\end{definition}

\begin{figure}[t]
    \centering
    \includegraphics[width = \linewidth]{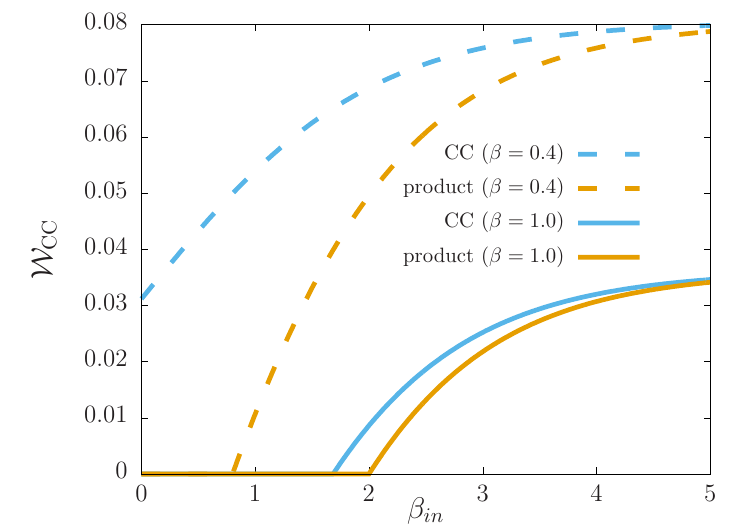}
    \caption{Daemonic ergotropy from the output of thermalizing maps in the case of classically controlled target and control. It can be seen that  at inverse maps' temperature $\beta=0.4$ the temperature bound is vanished while at $\beta=0.1$ is only shifted.}
    \label{fig:ErgotropyCC}
\end{figure}

Now, we report the result for the maximal SWITCH-ergotropy under classical priori correlations with respect to $N=2$ thermalizing channels.

\begin{theorem}\label{theor:ErgCC}
    The maximal SWITCH-ergotropy under classical correlations with respect to two thermalizing channels $\mathcal{T}_\beta$ is provided by the optimal state:
        \begin{equation}
    \rho_{SC}^{\mathrm{opt}} = \left\{
 \begin{array}{lc}
 \frac{1}{Z_{S,\beta_{\rm in}}} \Bigl( \ket{0+}\bra{0+} + e^{-\beta_{\rm in}} \ket{1-}\bra{1-} \Bigr), & \text{for} ~\beta_{\rm in} > 0  \\
 \frac{1}{2} \Bigl( \ket{++}\bra{++} + \ket{--}\bra{--} \Bigr), & \text{for} ~\beta_{\rm in} = 0 
\end{array}\right. 
    \end{equation}
and is given by:
\begin{equation}\label{eq:ErgotropyClassCorr}
    \mathcal{W}_{\mathrm{CC}} = \left\{
 \begin{array}{lc}
 \frac{\operatorname{max}\{ 0, e^{-2\beta} - e^{-\beta_{\rm in}} + 2e^{-(2\beta + \beta_{\rm in})}\}}{2Z_{S,\beta}^2Z_{S,\beta_{\rm in}}}, & \text{for} ~\beta_{\rm in} > 0  \\
 \frac{1}{2}\tanh\Bigl(\frac{\beta}{2}\Bigr)\Bigl( \sqrt{1 + \frac{1}{4}\sinh^{-2}(\beta)} - 1\Bigr), & \text{for} ~\beta_{\rm in} = 0 
\end{array}\right. .
    \end{equation}
    $\mathcal{W}_{\mathrm{CC}}$ is purely incoherent for $\beta_{\rm in} > 0$ and purely coherent for $\beta_{\rm in} = 0$.
    \begin{proof}
        See Appendix~\ref{appCC}.
    \end{proof}
\end{theorem}

As it can be witnessed from~(\ref{eq:ErgotropyClassCorr}), for $\beta_{\rm in} > 0$, allowing for initial classical correlations between work storage and control alters the temperature bound~(\ref{eq:TempLimit}) for the non-zero ergotropy.
\begin{corollary}
    If $\beta_{\rm in} > 0$, the maximal SWITCH-ergotropy~\eqref{eq:Erg2CCgen} under classical correlations is non-zero under condition
    \begin{equation}\label{eq:TempLimitClassCorr}
    \ln(e^{\beta_{\rm in}} + 2) > 2\beta.
    \end{equation}
    \begin{proof}
        For $\beta_{\rm in} > 0$, the condition $\mathcal{W}_{\mathrm{CC}} > 0$ leads to a condition $e^{-2\beta} - e^{-\beta_{\rm in}} + 2e^{-(2\beta + \beta_{\rm in})} > 0$, from which follows the temperature bound~(\ref{eq:TempLimitClassCorr}).
    \end{proof}
\end{corollary}
This means that initial classical correlations improve work extraction from the quantum SWITCH by shifting the temperature bound towards smaller $\beta_{\rm in}$ (see Fig.~\ref{fig:ErgotropyCC}). Moreover, it can easily seen that the left-hand side of~(\ref{eq:TempLimitClassCorr}) is bounded from below by $\ln(3)$ corresponding to the infinite initial temperature (i.e., $\beta_{\rm in} = 0$). Hence, the temperature $\tilde{T} = \frac{2}{\ln(3)}$ of the channels can be seen as a critical value, above which the bound~(\ref{eq:TempLimitClassCorr}) becomes trivial for any $T_{\rm in}$, while in the case of initially uncorrelated $S$ and $C$ this critical value is infinite. Indeed, for $T > \tilde{T}$, the ergotropy~(\ref{eq:ErgotropyClassCorr}) is activated in the entire range of initial temperatures $T_{\rm in}$, so that the temperature bound~(\ref{eq:TempLimit}) is overcome completely.

\begin{figure}[t!]
    \centering
    \includegraphics[width = \linewidth]{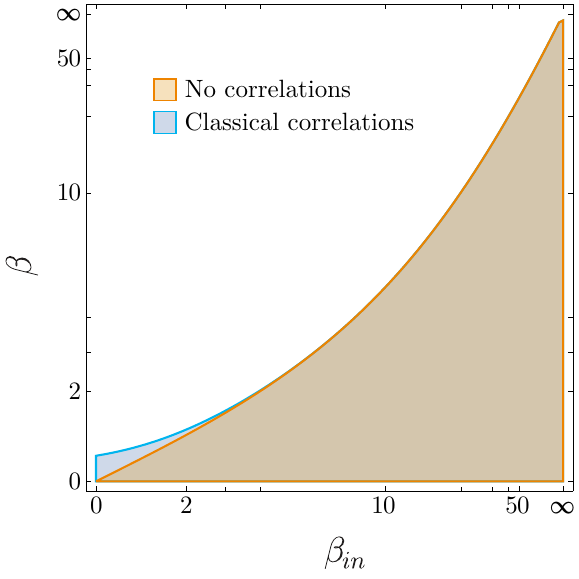}
    \caption{Comparison of regions of non-zero ergotropy in the space of $(\beta_{\rm in}, \beta)$-pairs for the initially uncorrelated and classically correlated target and control.}
    \label{fig:NonZeroErgotropyCC}
\end{figure}

On the other hand, for $T \leq \tilde{T}$, there still exists a finite range $\beta_{\rm in} \in (0,\ln (e^{2\beta} -2)]$, where the output state of $S$ becomes thermal and stores no work to be extracted. Therefore, we can conclude that prior classical correlations allow one to beat the temperature bound~(\ref{eq:TempLimit}) only partially. Indeed, quantum SWITCH assisted by initial classical correlations between $S$ and $C$ improves work extraction only in a tiny region of high temperatures $T_{\rm in}$ of initial state of the work medium and $T$ of the channels (i.e., low $\beta_{\rm in}$ and $\beta$, see Fig.~\ref{fig:NonZeroErgotropyCC}).

\subsection{Does assistance by prior quantum correlations perform better?}

Now, let us assume the quantum nature of the initial correlations between the work medium and control and question the advantages it can offer, particularly whether both temperature bounds~\eqref{eq:TempLimit} and \eqref{eq:TempLimitClassCorr} can be overcome in this setting. We focus on the maximal SWITCH-ergotropy under quantum correlations with respect to initial separable states with discord and initial entangled states, respectively.

\subsubsection{Quantum SWITCH with initial quantum discord between work storage and control system}
\label{sec:Discord}

Here we provide a calculation of daemonic ergotropy stored in the working medium after the action of the quantum SWITCH under a condition of its joint state with the control to be separable,
\begin{equation}
    \rho_{SC} = \sum_i p_i \rho_i \otimes \omega_i,
    \label{eq:sepstate}
\end{equation}
so that it can carry non-zero quantum discord. It has to satisfy the local thermality constraint~(\ref{eq:LocalThermality}) guaranteeing local passivity. To address this issue, we question an optimal separable state~(\ref{eq:sepstate}) that admits extraction of maximal daemonic ergotropy from the target system after the action of the quantum SWITCH of two thermalizing channels $\mathcal{T}_\beta$.

\begin{definition}
    The maximal SWITCH-ergotropy under initial quantum discord is the SWITCH-ergotropy with respect to two thermalizing processes $\mathcal{T}_\beta$ optimized over all initial states~(\ref{eq:sepstate}) carrying quantum discord:
    \begin{equation}
        \mathcal{W}_{\rm D} := \max_{\substack{\{p_i, \rho_i, \omega_i\}\\ \sum_i p_i \rho_i = \tau_{\beta_{\mathrm{in}}} }}  W_{\mathcal{S}}^{N=2}\Big(\sum_i p_i \rho_i \otimes \omega_i\Big),
    \end{equation}
    under the condition~(\ref{eq:LocalThermality}) of local thermality.
\end{definition}

Performing optimization of the overall daemonic ergotropy (See Appendix~\ref{appDiscord} for detailed calculations), we find two regimes of temperature pairs $(\beta, \beta_{\rm in})$, which feature different optimal initial states $\rho_{SC}$ and are separated by the temperature bound~(\ref{eq:TempLimit}).

\begin{theorem}\label{theor:ErgDiscord}
    The maximal SWITCH-ergotropy under quantum discord with respect to two thermalizing channels $\mathcal{T}_\beta$ is provided by the optimal state:
        \begin{equation}\label{eq:OptStateCC}
    \rho_{SC}^{\mathrm{opt}} = \left\{
 \begin{array}{lc}
 \frac{1}{Z_{S,\beta_{\rm in}}} \Bigl( \ket{0+}\bra{0+} + e^{-\beta_{\rm in}} \ket{1-}\bra{1-} \Bigr), & \text{for} ~\beta_{\rm in} \geq 2\beta  \\
 \frac{1}{2} \Bigl( \ket{\mu +}\bra{\mu +} + \ket{\nu -}\bra{\nu -} \Bigr), & \text{for} ~\beta_{\rm in} \leq 2\beta 
\end{array}\right. ,
    \end{equation}
where $|\mu/\nu\rangle = \frac{1}{\sqrt{Z_{S, \beta_{\rm in}}}} \Bigl(|0\rangle \pm e^{-\frac{\beta_{\rm in}}{2}} |1\rangle \Bigr)$, and is given by:
\begin{widetext}
\begin{equation}\label{eq:ErgDiscord}
    \mathcal{W}_{\mathrm{D}} = \left\{
     \begin{array}{lc}
        \frac{1}{2}\tanh\Bigl(\frac{\beta}{2}\Bigr)\Bigl( \cosh\bigl(\beta + \frac{\beta_{\rm in}}{2}\bigr)\xi(\beta, \beta_{\rm in}) - 1\Bigr), & \text{for} ~\beta_{\rm in} \geq 2\beta  \\
        \frac{1}{2}\tanh\Bigl(\frac{\beta}{2}\Bigr)\Bigl( \sqrt{1 + \xi^{2}(\beta, \beta_{\rm in})} - 1\Bigr), & \text{for} ~\beta_{\rm in} \leq 2\beta 
    \end{array}\right. ,
\end{equation}
\end{widetext}
where $\xi(\beta, \beta_{\rm in}) = \Bigl(2\sinh(\beta)\cosh\bigl(\frac{\beta_{\rm in}}{2}\bigr)\Bigr)^{-1}$. $\mathcal{W}_{\mathrm{D}}$ is purely incoherent for $\beta_{\rm in} > 2\beta$ and purely coherent for $\beta_{\rm in} < 2\beta$.
\begin{proof}
    See Appendix~\ref{appDiscord}.
\end{proof}
\end{theorem}

\begin{figure}[t]
    \centering
    \includegraphics[width = \linewidth]{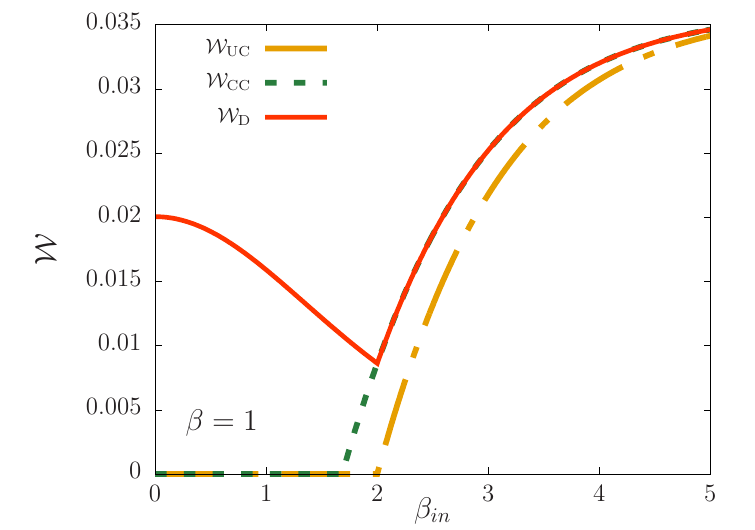}
    \caption{Maximal work $\mathcal{W}_{\rm D}$ that can be extracted from the work storage $S$ under initial quantum discord after the action of thermalizing channels at inverse temperature $\beta=1$ controlled by the quantum SWITCH depending on the temperature of the initial thermal state $\tau_{\beta_{\rm in}}$ (solid red line). It is compared with the maximal SWITCH-ergotropies without a priori correlations~(\ref{eq:2Erg}) and under classical correlations~\eqref{eq:ErgotropyClassCorr} (dot-dashed orange line and green dashed line, respectively).}
    \label{fig:ErgotropyQC}
\end{figure}

Theorem~\ref{theor:ErgDiscord} demonstrates that pre-shared quantum correlations between the work storage $S$ and control qubit $C$ in the form of discord reveal two regions of temperatures $(\beta_{\mathrm{in}}, \beta)$ divided by the temperature bound~(\ref{eq:TempLimit}). We start with the region of $\beta_{\rm in}$ violating the temperature bound~(\ref{eq:TempLimit}), which does not allow one to extract work from $S$ after action of the quantum SWITCH without initial correlations with $C$. The corresponding SWITCH-ergotropy is non-zero for almost any ($\beta_{\rm in}, \beta$)-pair, except for extreme cases of infinite inverse temperatures. Interestingly, it is constituted only by the coherent counterpart, so that the work can be extracted from $S$ only by consuming quantum coherence of its state. This stands in stark contrast with the gain in ergotropy with prior classical correlations (provided in Theorem~\ref{theor:ErgCC}) which has exclusively incoherent nature and is non-zero only in a certain region of temperature pairs ($\beta_{\rm in}, \beta$). Moreover, in contrast to the settings with initially uncorrelated or classically correlated $S$ and $C$, the SWITCH-ergotropy under initial quantum discord increases for higher input temperatures $T_{\rm in}$ (i.e., while decreasing $\beta_{\rm in}$, see Fig.~\ref{fig:ErgotropyQC}).

On the other hand, for temperatures satisfying the temperature bound~(\ref{eq:TempLimit}), we find an enhancement of SWITCH-ergotropy compared with one from the quantum SWITCH with initially uncorrelated $S$ and $C$. Interestingly, in this case, the resulting SWITCH-ergotropy has only incoherent counterpart and can be rewritten as
\begin{eqnarray}
    \nonumber \mathcal{W}_{\rm D} &=& \frac{1}{2Z_{S,\beta}^2Z_{S,\beta_{\rm in}}} \Bigl(e^{-2\beta} - e^{-\beta_{\rm in}} + 2e^{-(2\beta + \beta_{\rm in})}\Bigr), \\
    && \text{for} ~\beta_{\rm in} \geq 2\beta,
\end{eqnarray}
hence, coinciding with the SWITCH-ergotropy~(\ref{eq:ErgotropyClassCorr}) yielded in the scenario with prior classical correlations between $S$ and $C$. Indeed, the sharp change in the optimal configuration at $T_{\rm in} = \frac{T}{2}$ changes the nature of the resulting ergotropy from coherent one for higher input temperatures $(T_{\rm in} \geq \frac{T}{2})$ to incoherent one for lower input temperatures $(T_{\rm in} < \frac{T}{2})$. This results in a non-monotonic (yet continuous) behaviour of the maximal work~(\ref{eq:ErgDiscord}) that can be extracted from $S$, which decreases for $T_{\rm in} \geq \frac{T}{2}$ violating~(\ref{eq:TempLimit}) and increases for $T_{\rm in} < \frac{T}{2}$ satisfying~(\ref{eq:TempLimit}) while decreasing $T_{\rm in}$ (i.e., increasing $\beta_{\rm in}$).

\subsubsection{Prior entanglement is not more beneficial}

\begin{figure}[t]
    \centering
    \includegraphics[width = \linewidth]{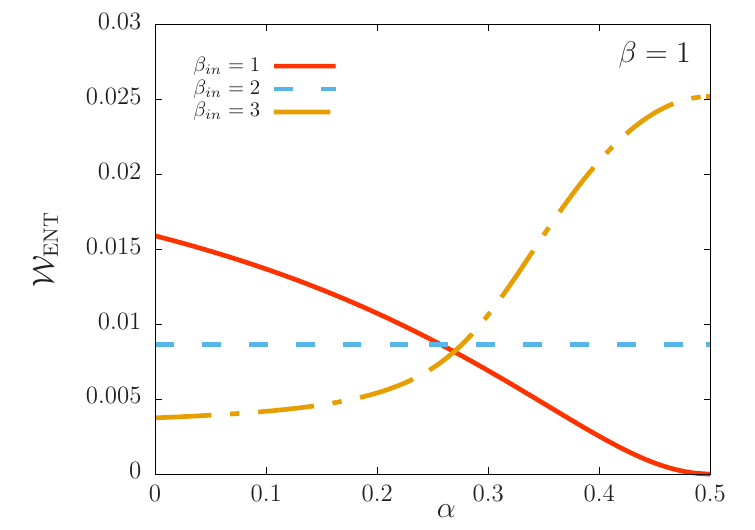}
    \caption{SWITCH-ergotropy depending on the purification~(\ref{eq:purification}) of the initial thermal state $\tau_{\beta_{\rm in}}$. For $\beta_{\rm in} \leq 2\beta$, the optimal purification is provided by the choices $\alpha = 0$ and $\alpha = 1$. For $\beta_{\rm in} \geq 2\beta$, the optimal purification is yielded by $\alpha = \frac12$.}
    \label{fig:EntanglementPurification}
\end{figure}

Now we focus our analysis on the case when $S$ and $C$ initially share quantum entanglement under the constraint of initial local thermality of the target system $S$ via the condition~(\ref{eq:LocalThermality}). This means that the initial joint state of $S$ and $C$ can be seen as a purification of the thermal state of the target system.
\begin{definition}
    A purification $(\alpha, \phi)$ of the initial thermal state of the target system $S$ with respect to the control system $C$ is a joint entangled state
\begin{equation}\label{eq:purification}
|\Phi_{\alpha, \phi}(\beta_{\rm in})\rangle = \frac{1}{\sqrt{Z_{S,\beta_{\rm in}}}} \Bigl( |0\psi_{\alpha, \phi}\rangle + e^{-\frac{\beta_{\rm in}}{2}}|1\psi_{\alpha, \phi}^{\perp}\rangle \Bigr),
\end{equation}
where the energetic eigenstates of $S$ are entangled with the states 
\begin{eqnarray}
    |\psi_{\alpha, \phi}\rangle &=& \sqrt{\alpha}|0\rangle + e^{i \phi} \sqrt{1-\alpha} |1\rangle, \\
    |\psi_{\alpha, \phi}^\perp\rangle &=& e^{-i \phi} \sqrt{1-\alpha}|0\rangle -  \sqrt{\alpha} |1\rangle,
\end{eqnarray}
of $C$ constituting an arbitrary orthonormal basis with $\alpha \in [0,1]$ and $\phi \in [0,2\pi]$\footnote{In the case of $\{|\psi\rangle, |\psi^\perp\rangle\}$ representing energetic eigenstates of $C$, $|\Phi_{\alpha, \phi}(\beta_{\rm in})\rangle$ is an example of states known as thermofield double states~\cite{Takahashi1996, Cottrell2019}.}.
\end{definition}
This means that the joint state satisfying~(\ref{eq:LocalThermality}) corresponding to a purification $(\alpha, \phi)$ is given by $\rho_{SC} = |\Phi_{\alpha, \phi}(\beta_{\rm in})\rangle \langle \Phi_{\alpha, \phi}(\beta_{\rm in})|$. Similarly to the case of initial discord, our figure of merit is the maximal daemonic ergotropy that can be extracted from $S$ after the action of the quantum SWITCH.
\begin{definition}
    The maximal SWITCH-ergotropy under initial quantum entanglement is the SWITCH-ergotropy with respect to two thermalizing processes $\mathcal{T}_\beta$ optimized over all initial entangled states~(\ref{eq:purification}):
    \begin{equation}\label{eq:MaxWorkQC}
    \mathcal{W}_{\rm ENT} := \max_{\alpha, \phi} W_{\mathcal{S}} \Big(|\Phi_{\alpha, \phi}(\beta_{\rm in})\rangle \langle \Phi_{\alpha, \phi}(\beta_{\rm in})|\Big).
    \end{equation}
\end{definition}
Interestingly, as in the case of initial discord, the optimal purification $(\alpha_{\rm{opt}}, \phi_{\rm {opt}})$ leading to the maximal daemonic ergotropy depends crucially on relation between the temperatures $\beta_{\rm in}$ of the initial thermal state and $\beta$ of the maps.

\begin{theorem}\label{theor:opt_ent}
    The maximal SWITCH-ergotropy under initial quantum entanglement with respect to two thermalizing channels $\mathcal{T}_\beta$ is provided by the optimal purification
\begin{eqnarray}
\alpha_{\rm{opt}}(\beta_{\rm in}, \beta) &=&  \left\{
 \begin{array}{cc}
 0 \text{ or } 1, & \text{for} ~\beta_{\rm in} \leq 2 \beta  \\
 \frac12, & \text{for} ~\beta_{\rm in} \geq 2 \beta
\end{array}\right.,
\end{eqnarray}
with arbitrary $\phi$ and is given by:
\begin{widetext}
\begin{equation}\label{eq:ErgQC}
    \mathcal{W}_{\mathrm{ENT}} = \left\{
    \begin{array}{lc}
        \frac{1}{2}\tanh\Bigl(\frac{\beta}{2}\Bigr)\Bigl( \cosh\bigl(\beta + \frac{\beta_{\rm in}}{2}\bigr)\xi(\beta, \beta_{\rm in}) - 1\Bigr), & \text{for} ~\beta_{\rm in} \geq 2\beta  \\
        \frac{1}{2}\tanh\Bigl(\frac{\beta}{2}\Bigr)\Bigl( \sqrt{1 + \xi^{2}(\beta, \beta_{\rm in})} - 1\Bigr), & \text{for} ~\beta_{\rm in} \leq 2\beta 
    \end{array}\right. ,
    \end{equation}
\end{widetext}
where $\xi(\beta, \beta_{\rm in}) = \Bigl(2\sinh(\beta)\cosh\bigl(\frac{\beta_{\rm in}}{2}\bigr)\Bigr)^{-1}$. $\mathcal{W}_{\mathrm{ENT}}$ is purely incoherent for $\beta_{\rm in} > 2\beta$ and purely coherent for $\beta_{\rm in} < 2\beta$.
\begin{proof}
See Appendix~\ref{appQC}.
\end{proof}

\end{theorem}

Furthermore, note that the SWITCH-ergotropy~(\ref{eq:ErgQC}) under initial quantum entanglement, similarly to the case of initial quantum discord, has a non-monotonic behaviour (see Fig.~\ref{fig:EntanglementPurification}). Moreover, comparing it with~(\ref{eq:ErgDiscord}), we find that initial entanglement\footnote{The reason for choosing a purified thermal state \eqref{eq:purification} is the fact that this is the state with maximal entanglement with a local thermal marginal, where its entanglement entropy is equal to the von Neumann entropy of the local thermal state, providing optimal SWITCH-ergotropy (see \ref{app:mixedQC}).} does not perform better than initial quantum discord in terms of maximal daemonic ergotropy of the quantum SWITCH and conclude with the following Corollary.

\begin{corollary}
    The maximal SWITCH-ergotropy under initial quantum entanglement coincides with one under initial quantum discord:
    \begin{equation}
        \mathcal{W}_{\rm ENT} = \mathcal{W}_{\rm D}.
    \end{equation}
\begin{proof}
    The proof follows immediately by comparing the SWITCH-ergotropies~(\ref{eq:ErgDiscord}) and~(\ref{eq:ErgQC}) under initial quantum discord and quantum entanglement, respectively.
\end{proof}
\end{corollary}

\section{Conclusions}\label{sec:conclusions}
Thermalization phenomena can be regarded as processes that degrade a thermodynamic resource carried by a physical system. However, in the quantum domain, a possibility of putting several thermalization processes into an indefinite causal order, e.g., via the quantum SWITCH, challenges this view. Indeed, an apparently thermodynamically useless scenario of a qubit prepared in a thermal state and undergoing action of a sequence of thermalization processes can become a work resource if their order is controlled by the quantum SWITCH~\cite{Simonov2022}. Nevertheless, quantum SWITCH exhibits a constraint on initial configuration of temperatures of the system and the channels representing thermalization processes. Indeed, a non-zero work stored a posteriori in the qubit requires the latter to be initially colder than half-temperature of the thermalizing channels in the considered scenario.

In this work, we have addressed several approaches to overcome this temperature bound. At first, one can question its validity in a scenario of arbitrary number of thermalization processes executed via the quantum SWITCH. Given a qubit system satisfying the temperature bound, it stores increasingly more work while increasing the number of the thermalizing channels. However, even in this setting, the regions of temperatures violating the considered bound remain thermodynamically useless.

On the other hand, the quantum SWITCH does not necessarily require the control system to share no prior correlations with the system of interest as in earlier works (c.f.~\cite{Guha2023}) that induces superposition of causal order. Indeed, the quantum SWITCH can support a more general scenario of a qubit which is initially in a thermal state only locally (i.e., its marginal state is thermal) while being correlated with the control. Interestingly, we found that, in contrast to multiple thermalization processes, assistance of the quantum SWITCH by prior correlations can shift and even completely break the temperature bound. Crucially, the qubit system does not store any work even being supported by these correlations if the thermalization processes are realized in a well-defined order or in one chosen randomly.

While quantum systems can support a wide spectrum of correlations between them, we have analyzed several settings of the qubit work medium and control sharing prior correlations of different nature and explored their impact on work stored in the former after action of the quantum SWITCH. At first, we have examined a scenario of a priori classically correlated work medium and control. In this case, we have found a wider domain of initial temperatures, where a non-zero work extraction is allowed, compared with one produced by the scenario of initially uncorrelated work medium and control~\cite{Simonov2022}. Indeed, there exists a critical value of temperature of thermalizing channels, above which the qubit system appears to store work regardless of its initial temperature. On the other hand, below the critical temperature, there still exists a constraint on initial temperatures of the work medium. Nevertheless, it includes a greater domain of initial temperatures compared with the temperature bound from the uncorrelated scenario and can be regarded as a shift of the latter.

In the realm of quantum correlations, we have focused on two scenarios of work medium and control assisted by prior quantum discord and quantum entanglement, respectively. Crucially, initial quantum entanglement does not achieve more work stored in the qubit system after action of the quantum SWITCH compared with the case of prior quantum discord. In stark contrast to initial classical correlations, a priori quantum correlations yield non-zero work after action of the quantum SWITCH regardless of initial temperatures of the work medium and the thermalizing channels. Moreover, we found that there exist two domains of initial temperatures, where the work stored in the qubit system is of different nature: exclusively incoherent (work can be extracted by exchanging the energy levels' populations of the state) and coherent (work can be extracted by consuming quantum coherence of the state) one. The bound separating these domains appears to coincide with the temperature bound from the uncorrelated scenario. Crucially, in the region of temperatures forbidden by the latter, assistance by prior entanglement yields non-zero work which is not achievable neither with uncorrelated nor classical correlations. On the other hand, in the region of temperatures satisfying the temperature bound from the uncorrelated scenario, assistance by prior entanglement does not achieve more work than one by prior classical correlations. Quantification of the thermodynamic cost of generating correlations in the quantum realm has been discussed in the literature \cite{Huber2015}. In general, the costs of generating classical correlations $\mathcal{C}_{\text{CC}}$, discord $\mathcal{C}_{\text{D}}$, and entanglement $\mathcal{C}_{\text{ENT}}$, respectively, satisfy the bounds $\mathcal{C}_{\text{CC}} \leq \mathcal{C}_{\text{D}}\leq \mathcal{C}_{\text{ENT}}$. Although the finite cost of creating such correlations did not enter our analysis, we found that the task of work extraction after action of thermalizing channels controlled by the quantum SWITCH establishes a strict outperformance of a priori classical correlations by prior quantum discord, while the latter provides as good assistance as prior quantum entanglement. In particular, this means that the higher thermodynamic cost for generating entanglement does not translate into a larger ergotropy gain. This finding represents an interesting property of the quantum SWITCH and provides an example of a situation where quantum discord is strictly better than classical correlations, but entanglement cannot outperform quantum discord.

The obtained results suggest that the quantum SWITCH can lie in the basis of a new class of thermodynamic protocols able to activate thermodynamically free resources (e.g., charge a quantum battery). Taking into account that the quantum SWITCH is a supermap realizing an indefinite causal order of the maps fed into it (in our case, resourceless thermalizing channels), it is necessary to question the role of causal non-separability in thermodynamics. This can be verified via existing implementations of the quantum SWITCH in various experimental platforms, including photonic setups \cite{Rozema2024} and NMR devices \cite{Vieira2023}. Notably, the quantum SWITCH involving thermodynamically free channels has been successfully realized in several experimental studies \cite{Cao2022, Nie2022_Exp, Xi2024, Tang2024}. Moreover, the reverse task, namely, the charging of a work storage device has been addressed as well. It has been demonstrated that this process can benefit from indefinite causal order of operations, yielding enhanced performance relative to conventional charging protocols \cite{Zhu2023}. Therefore, an interesting direction of future research is to examine causal non-separability as a hypothetical thermodynamic resource. This requires a shift in paradigm of quantum thermodynamics towards a higher-order one. One of the options would be consideration of thermodynamics as a resource theory and development of one based on higher-order maps as the figure of merit. Hence, we hope that our work will stimulate the research in this direction.

\section*{Acknowledgements}
We acknowledge support of the Hong Kong Research Grant Council through grants 17326619 and 17300920, and through the Senior Research Fellowship Scheme via SRFS2021-7S02, the John Templeton Foundation through grant  62312, The Quantum Information Structure of Spacetime (qiss.fr). Research at the Perimeter Institute is supported by the Government of Canada through the Department of Innovation, Science and Economic Development Canada and by the Province of Ontario through the Ministry of Research, Innovation and Science. The opinions expressed in this publication are those of the authors and do not necessarily reflect the views of the John Templeton Foundation. This research was funded in whole or in part by the Austrian Science Fund (FWF) 10.55776/PAT4559623. For open access purposes, the author has applied a CC BY public copyright license to any author-accepted manuscript version arising from this submission.

\bibliography{main}

\onecolumngrid
\appendix

\section{Proofs of Theorem~\ref{theor:ErgUC2} and Theorem~\ref{theor:ErgUCN}}
\label{appNC}

In this Appendix, we review the results of~\cite{Simonov2022} for the quantum SWITCH for two thermalizing channels and generalize them to the case of $N$ thermalizing channels.

\subsection{Proof of Theorem~\ref{theor:ErgUC2}}
\label{appUC2}

For $N=2$ channels, the action of the quantum SWITCH is provided by~(\ref{uncorrswitch}). Before to proceed with it, we prove the following Lemma that facilitates the calculations in the case of identical controlled channels.

\begin{lemma}\label{lemma:app:IdentChan}
    Given $2$ identical channels $\mathcal{E}[\cdot]$, the action of the quantum SWITCH controlling their causal order can be written as:
    \begin{equation}\label{eq:app:IdentChanGen}
    \big[ \map S (\map{E}, \map{E}) \big]  (\rho_{\rm in} \otimes \omega )  = \frac{1}{4} \sum_{ij} \Bigl( \{E_i, E_j\} \rho_{\rm in} \{E_i, E_j\}^\dagger \otimes \omega + [E_i, E_j] \rho_{\rm in} [E_i, E_j]^\dagger \otimes Z\omega Z \Bigr),
    \end{equation}
    where $\{E_i\}_i$ is the Kraus decomposition of $\mathcal{E}$.
    \begin{proof}
        The proof follows immediately by observation that $[E_i, E_j] \rho_{\rm in} \{E_i, E_j\}^\dagger = (\{E_i, E_j\} \rho_{\rm in} [E_i, E_j]^\dagger)^\dagger$ and:
        \begin{align}
            \nonumber \sum_{ij} \{E_i, E_j\} \rho_{\rm in} [E_i, E_j]^\dagger &= \sum_{ij} \Bigl( E_i E_j \rho_{\rm in} (E_i E_j)^\dagger + E_j E_i \rho_{\rm in} (E_i E_j)^\dagger - E_i E_j \rho_{\rm in} (E_j E_i)^\dagger - E_j E_i \rho_{\rm in} (E_j E_i)^\dagger \Bigr) \\
            \nonumber &= \sum_{ij} E_i E_j \rho_{\rm in} (E_i E_j)^\dagger + \sum_{ij} E_j E_i \rho_{\rm in} (E_i E_j)^\dagger - \sum_{ij} E_j E_i \rho_{\rm in} (E_i E_j)^\dagger - \sum_{ij} E_i E_j \rho_{\rm in} (E_i E_j)^\dagger \\
            &= 0.
        \end{align}
    \end{proof}
\end{lemma}

Now, taking the initial state of $S$ as a thermal state $\tau_{\beta_{\rm in}}$ and two identical thermalizing channels $\mathcal{T}_\beta$, we question first the optimal control state $\omega$ that maximizes the SWITCH-ergotropy. 
\begin{proposition}\label{app:prop:OptStateUC}
     The SWITCH-ergotropy~(\ref{def:SwitchErg}) with $\rho_{SC} = \tau_{\beta_{\rm in}} \otimes \omega$ is maximal for $\omega = |+\rangle\langle +|$.
     \begin{proof}
         First, we find the output of the quantum SWITCH by applying Lemma~\ref{lemma:app:IdentChan}:
    \begin{equation}\label{eq:app:IdentChan}
    \big[ \map S (\mathcal{T}_\beta, \mathcal{T}_\beta) \big]  (\tau_{\beta_{\rm in}} \otimes \omega ) = \sigma_+ \otimes \omega + \sigma_- \otimes Z\omega Z,
    \end{equation}
where 
\begin{align}
\nonumber \sigma_+ &= \frac{1}{4} \sum_{iji'j'} \{T_{\beta, ij}, T_{\beta, i'j'}\} \tau_{\beta_{\rm in}} \{T_{\beta, ij}, T_{\beta, i'j'}\}^\dagger, \\
\nonumber \sigma_- &= \frac{1}{4} \sum_{iji'j'} [T_{\beta, ij}, T_{\beta, i'j'}] \tau_{\beta_{\rm in}} [T_{\beta, ij}, T_{\beta, i'j'}]^\dagger.
\end{align}
The maximal SWITCH-ergotropy is given then by optimization of its daemonic ergotropy over all states $\omega$. Applying the definition of daemonic ergotropy~(\ref{DaemonicErg}), we can write the maximal SWITCH-ergotopy as:
\begin{equation}
    \mathcal{W}_{\rm UC}^{N=2} = \max_{\omega, |\psi\rangle} \Biggl( W\Bigl(\langle \psi | \omega | \psi \rangle \sigma_+ + \langle \psi | Z\omega Z | \psi \rangle \sigma_- \Bigr) + W\Bigl(\langle \psi^\perp | \omega | \psi^\perp \rangle \sigma_+ + \langle \psi^\perp | Z\omega Z | \psi^\perp \rangle \sigma_- \Bigr)\Biggr),
\end{equation}
where $\{|\psi\rangle, |\psi^\perp\rangle\}$ form the measurement basis, and the identity $p_i W(\rho_{S|i}) = W(p_i \rho_{S|i})$ is taken into account in~(\ref{DaemonicErg}). On the other hand, we can rewrite $\sigma_+ + \sigma_- = (\mathcal{T}_\beta \circ \mathcal{T}_\beta)[\tau_{\beta_{\rm in}}] = \tau_\beta$, hence, obtaining
\begin{eqnarray}
    \nonumber \mathcal{W}_{\rm UC}^{N=2} &=& \frac{1}{2} \max_{\omega, |\psi\rangle} \Biggl( W\Bigl(\langle \psi | (\omega + Z\omega Z) | \psi \rangle \tau_\beta + \langle \psi | (\omega - Z\omega Z) | \psi \rangle G \Bigr) \\
    &+& W\Bigl(\langle \psi^\perp | (\omega + Z\omega Z) | \psi^\perp \rangle \tau_\beta + \langle \psi^\perp | (\omega - Z\omega Z) | \psi^\perp \rangle G \Bigr)\Biggr),
\end{eqnarray}
where $G \equiv \sigma_+ - \sigma_-$. Taking into account that $\omega + Z\omega Z = \mathds{1} + \langle Z \rangle_\omega Z$ and $\omega - Z\omega Z = \langle X \rangle _\omega X + \langle Y \rangle_\omega Y$, where $\langle A \rangle_\omega = \operatorname{Tr}[A\omega] \equiv \omega_A$ with $\omega_A$ being a component of the corresponding Bloch vector, we find
\begin{align}\label{app:eq:2erg}
    \nonumber \mathcal{W}_{\rm UC}^{N=2} &= \frac{1}{2}\max_{\omega, |\psi\rangle} \Biggl( W\Bigl( (1+\tilde{p}(\psi, \omega)) \tau_\beta + \lambda (\psi, \omega) G \Bigr) \\
    &+ W\Bigl( (1 - \tilde{p}(\psi, \omega)) \tau_\beta - \lambda(\psi, \omega) G \Bigr)\Biggr),
\end{align}
where $\tilde{p}(\psi, \omega) = \langle Z \rangle_\psi \langle Z \rangle_\omega$ and $\lambda(\psi, \omega) = \langle X \rangle_\psi \langle X \rangle_\omega + \langle Y \rangle_\psi \langle Y \rangle_\omega$. As $\tau_\beta$ is a thermal state, hence, carrying zero ergotropy, the only possible source of non-zero ergotropy in~(\ref{app:eq:2erg}) are the terms proportional to $G$. Therefore, optimization of the SWITCH-ergotropy~(\ref{app:eq:2erg}) is achieved by maximal value of $\lambda(\psi, \omega)$, i.e., by the states $|\psi\rangle$ and $\omega$ with $|\langle X \rangle_\psi| = |\langle X \rangle_\omega| = 1$ or $|\langle Y \rangle_\psi| = |\langle Y \rangle_\omega| = 1$. This means that the optimal control state $\omega$ is pure and can be taken as one of the states $|\pm\rangle = \frac{|0\rangle \pm |1\rangle}{\sqrt{2}}$ and $|\pm i\rangle = \frac{|0\rangle \pm i |1\rangle}{\sqrt{2}}$. Without loss of generality, we choose $\omega = |+\rangle\langle +|$, hence the proof.
     \end{proof}
\end{proposition}

Finally, we can calculate the resulting maximal SWITCH-ergotropy, as shown in the following Proposition.
\begin{proposition}
The maximal SWITCH-ergotropy without a priori correlations is given by:
\begin{equation}\label{app:prop:eq:2Erg}
        \mathcal{W}_{\mathrm{UC}}^{N=2} = \frac{\operatorname{max}\{ 0, e^{-2\beta} - e^{-\beta_{\rm in}}\}}{2Z_{S,\beta}^2Z_{S,\beta_{\rm in}}},
    \end{equation}
\begin{proof}
Following~(\ref{eq:app:IdentChan}), we have to calculate the unnormalized states $\sigma_\pm$. For this purpose, we report the anticommutators of the Kraus operators~(\ref{eq:ThermalOp})
\[\arraycolsep=6.4pt\def\arraystretch{2.2}
\begin{array}{llll}
       \{T_{\beta, 00}, T_{\beta, 00} \} = \frac{2}{Z_\beta}  |0\rangle \langle 0 | & \{T_{\beta, 00}, T_{\beta, 01} \} = \frac{1}{Z_\beta}  |0\rangle \langle 1 | & \{T_{\beta, 00}, T_{\beta, 10} \} = \frac{e^{-\frac{\beta}{2}}}{Z_\beta} |1\rangle \langle 0 | & \{T_{\beta, 00}, T_{\beta, 11} \} = 0 \\
       \{T_{\beta, 01}, T_{\beta, 00} \} = \frac{1}{Z_\beta}  |0\rangle \langle 1 | & \{T_{\beta, 01}, T_{\beta, 01} \} = 0 & \{T_{\beta, 01}, T_{\beta, 10} \} = \frac{e^{-\frac{\beta}{2}}}{Z_\beta} \mathds{1} & \{T_{\beta, 01}, T_{\beta, 11} \} = \frac{e^{-\frac{\beta}{2}}}{Z_\beta} |0\rangle \langle 1 | \\
       \{T_{\beta, 10}, T_{\beta, 00} \} = \frac{e^{-\frac{\beta}{2}}}{Z_\beta} |1\rangle \langle 0 | & \{T_{\beta, 10}, T_{\beta, 01} \} = \frac{e^{-\frac{\beta}{2}}}{Z_\beta} \mathds{1} & \{T_{\beta, 10}, T_{\beta, 10} \} = 0 & \{T_{\beta, 10}, T_{\beta, 11} \} = \frac{e^{-\beta}}{Z_\beta} |1\rangle \langle 0 | \\
        \{T_{\beta, 11}, T_{\beta, 00} \} = 0 & \{T_{\beta, 11}, T_{\beta, 01} \} = \frac{e^{-\frac{\beta}{2}}}{Z_\beta} |0\rangle \langle 1 | & \{T_{\beta, 11}, T_{\beta, 10} \} = \frac{e^{-\beta}}{Z_\beta} |1\rangle \langle 0 | & \{T_{\beta, 11}, T_{\beta, 11} \} = \frac{2e^{-\beta}}{Z_\beta} |1\rangle \langle 1 |
\end{array}
\]
and their commutators,
\[\arraycolsep=6.4pt\def\arraystretch{2.2}
\begin{array}{llll}
       [T_{\beta, 00}, T_{\beta, 00} ] = 0 & [T_{\beta, 00}, T_{\beta, 01} ] = \frac{1}{Z_\beta}  |0\rangle \langle 1 | & [T_{\beta, 00}, T_{\beta, 10} ] = -\frac{e^{-\frac{\beta}{2}}}{Z_\beta} |1\rangle \langle 0 | & [T_{\beta, 00}, T_{\beta, 11} ] = 0 \\
       \lbrack T_{\beta, 01}, T_{\beta, 00} ] = -\frac{1}{Z_\beta}  |0\rangle \langle 1 | & [T_{\beta, 01}, T_{\beta, 01} ] = 0 & [T_{\beta, 01}, T_{\beta, 10} ] = \frac{e^{-\frac{\beta}{2}}}{Z_\beta} Z & [T_{\beta, 01}, T_{\beta, 11} ] = \frac{e^{-\frac{\beta}{2}}}{Z_\beta} |0\rangle \langle 1 | \\
       \lbrack T_{\beta, 10}, T_{\beta, 00} ] = \frac{e^{-\frac{\beta}{2}}}{Z_\beta} |1\rangle \langle 0 | & [T_{\beta, 10}, T_{\beta, 01} ] = -\frac{e^{-\frac{\beta}{2}}}{Z_\beta} Z & [T_{\beta, 10}, T_{\beta, 10} ] = 0 & [T_{\beta, 10}, T_{\beta, 11} ] = -\frac{e^{-\beta}}{Z_\beta} |1\rangle \langle 0 | \\
       \lbrack T_{\beta, 11}, T_{\beta, 00} ] = 0 & [T_{\beta, 11}, T_{\beta, 01} ] = -\frac{e^{-\frac{\beta}{2}}}{Z_\beta} |0\rangle \langle 1 | & [T_{\beta, 11}, T_{\beta, 10} ] = \frac{e^{-\beta}}{Z_\beta} |1\rangle \langle 0 | & [T_{\beta, 11}, T_{\beta, 11} ] = 0
\end{array}
\]
and obtain
\begin{eqnarray}
    \nonumber \sigma_\pm &=& \frac{1}{2Z_\beta^2}\Bigl[ |0\rangle\langle 0| \tau_{\beta_{\rm in}} |0\rangle\langle 0| \pm |0\rangle\langle 0| \tau_{\beta_{\rm in}} |0\rangle\langle 0| + |0\rangle\langle 1| \tau_{\beta_{\rm in}} |1\rangle\langle 0| \\
    \nonumber &&+ \; e^{-\beta} \Bigl( |0\rangle\langle 0| \tau_{\beta_{\rm in}} |0\rangle\langle 0| + |0\rangle\langle 1| \tau_{\beta_{\rm in}} |1\rangle\langle 0| + |1\rangle\langle 0| \tau_{\beta_{\rm in}} |0\rangle\langle 1| + |1\rangle\langle 1| \tau_{\beta_{\rm in}} |1\rangle\langle 1| \Bigr) \\
    \nonumber &&+ \; e^{-2\beta} \Bigl( |1\rangle\langle 0| \tau_{\beta_{\rm in}} |0\rangle\langle 1| + |1\rangle\langle 1| \tau_{\beta_{\rm in}} |1\rangle\langle 1| \pm |1\rangle\langle 1| \tau_{\beta_{\rm in}} |1\rangle\langle 1| \Bigr) \Bigr] \\
    \nonumber &=& \frac{1}{2Z_\beta^2}\Bigl[ |0\rangle\langle 0| \pm |0\rangle\langle 0| \tau_{\beta_{\rm in}} |0\rangle\langle 0| + e^{-\beta} \Bigl( |0\rangle\langle 0| + |1\rangle\langle 1| \Bigr) + e^{-2\beta} \Bigl( |1\rangle\langle 1| \pm |1\rangle\langle 1| \tau_{\beta_{\rm in}} |1\rangle\langle 1| \Bigr) \Bigr] \\
    \nonumber &=& \frac{1}{2Z_\beta}\Bigl[ |0\rangle\langle 0| + e^{-\beta} |1\rangle\langle 1| \pm \frac{1}{Z_\beta} \Bigl( |0\rangle\langle 0| \tau_{\beta_{\rm in}} |0\rangle\langle 0| + e^{-2\beta} |1\rangle\langle 1| \tau_{\beta_{\rm in}} |1\rangle\langle 1| \Bigr) \Bigr] \\
    &=& \frac{1}{2}[ \tau_\beta \pm \tau_\beta \tau_{\beta_{\rm in}} \tau_\beta ].
\end{eqnarray}
Taking into account that maximal SWITCH-ergotropy is achieved by $\omega = |+\rangle\langle +|$ and measurements in $|\pm\rangle$-basis, we obtain $\tilde{p}(\psi, \omega) = 0$ and $\lambda(\psi, \omega) = 1$ and applying~(\ref{app:eq:2erg}), we obtain
\begin{equation}
    \mathcal{W}_{\rm UC}^{N=2} = \frac{1}{2} W\Bigl( \tau_\beta + \tau_\beta \tau_{\beta_{\rm in}} \tau_\beta \Bigr) + \frac{1}{2} W\Bigl( \tau_\beta - \tau_\beta \tau_{\beta_{\rm in}} \tau_\beta \Bigr),
\end{equation}
Since the term $\tau_\beta \tau_{\beta_{\rm in}} \tau_\beta$ does not carry quantum coherence in the energetic basis, the corresponding ergotropy consists of \textit{incoherent} counterpart only, which can be given as~\cite{Simonov2022}
\begin{equation}\label{app:DaemIncoherErg}
    \mathcal{W}_{\rm UC}^{N=2} = \frac{1}{2}\operatorname{max}\{0, |\delta G| - |\delta \rho_{\rm def}|\},
\end{equation}
where $\delta G$ is the difference of the energetic populations of $G = \tau_\beta \tau_{\beta_{\rm in}} \tau_\beta$, and $\delta \rho_{\rm def}$ is the difference of the energetic populations of $\rho_{\rm def} = \tau_\beta$ of the output of causally separable combination of $\mathcal{T}_\beta$, i.e., $(\mathcal{T}_\beta \circ \mathcal{T}_\beta)$. Therefore, quantum SWITCH activates ergotropy
\begin{eqnarray}
     \mathcal{W}_{\rm UC}^{N=2} &=& \frac{1}{2Z_{S,\beta}^2 Z_{S,\beta_{\rm in}}}\operatorname{max}\Bigl\{ 0, 1 - e^{-(2\beta + \beta_{\rm in})} - Z_{S,\beta} Z_{S,\beta_{\rm in}} (1 - e^{-\beta})\Bigr\} \\
     &=& \frac{1}{2Z_{S,\beta}^2 Z_{S,\beta_{\rm in}}}\operatorname{max}\Bigl\{ 0, e^{-2\beta} - e^{-\beta_{\rm in}} \Bigr\},
\end{eqnarray}
hence the proof. 

\end{proof}
\end{proposition}

\subsection{Proof of Theorem~\ref{theor:ErgUCN}}
\label{appUCN}

In what follows, we generalize the above results to the case of arbitrary number $N$ of thermalizing channels. We start by the following three Lemmas that will simplify the proof of Theorem~\ref{theor:ErgDiscord}

\begin{lemma}\label{app:lemma:NUCoutput}
    Given an initial uncorrelated joint state $\rho_{SC} = \tau_{\beta_{\rm in}} \otimes \omega$, where $\omega$ is a state of the $N$-dimensional control system, the output of the quantum SWITCH of $N$ identical thermalizing channels $\mathcal{T}_\beta[\cdot]$ is given by:
    \begin{equation}\label{app:eq:NOutput}
        \big[\mathcal{S}(\mathcal{T}_{\beta}, ..., \mathcal{T}_{\beta})\big] (\tau_{\beta_{\rm in}} \otimes \omega) = \sigma_+ \otimes \omega + \sigma_- \otimes \tilde{\omega},
    \end{equation}
    where $\tilde{\omega}$ is a state obtained by inverting the sign of off-diagonal elements of $\omega$, and $\sigma_\pm = \frac{1}{2}(\tau_\beta \pm G)$ with $G = \tau_\beta \tau_{\beta_{\rm in}} \tau_\beta$.
    \begin{proof}
        For $N$ thermalizing channels, the output of the corresponding $N$-SWITCH can be given as~\cite{Nie2022}
        \begin{align}
            \nonumber \big[\mathcal{S}(\mathcal{T}_{\beta}, ..., \mathcal{T}_{\beta})\big] (\tau_{\beta_{\rm in}} \otimes \omega) &= \sum_{i_1, j_1, ..., i_N, j_N} \Biggl( \sum_i \mathcal{P}_{i}\Bigl(T_{\beta, i_1 j_1} ... T_{\beta, i_N j_N}\Bigr) \tau_{\beta_{\rm in}} \Bigl(\mathcal{P}_{i}\Bigl(T_{\beta, i_1 j_1} ... T_{\beta, i_N j_N}\Bigr)\Bigr)^\dagger \otimes \omega_{ii} |i\rangle\langle i| \\
            &+ \sum_{i\neq j} \mathcal{P}_{i}\Bigl(T_{\beta, i_1 j_1} ... T_{\beta, i_N j_N}\Bigr) \tau_{\beta_{\rm in}} \Bigl(\mathcal{P}_{j}\Bigl(T_{\beta, i_1 j_1} ... T_{\beta, i_N j_N}\Bigr)\Bigr)^\dagger \otimes \omega_{ij} |i\rangle \langle j| \Biggr),
        \end{align}
        where $\omega_{ij}$ are $ij$-elements of the state $\omega$. The action of Kraus operators in the first term is equivalent to the action of a composition of $N$ thermalizing channels and, therefore, outputs the thermal state $\tau_\beta$. On the other hand, as demonstrated in~\cite{Nie2022}, the action of Kraus operators in the second term can be given as:
        \begin{equation}
            \sum_{i_1, j_1, ..., i_N, j_N} \mathcal{P}_{i}\Bigl(T_{\beta, i_1 j_1} ... T_{\beta, i_N j_N}\Bigr) \tau_{\beta_{\rm in}} \Bigl(\mathcal{P}_{j \neq i}\Bigl(T_{\beta, i_1 j_1} ... T_{\beta, i_N j_N}\Bigr)\Bigr)^\dagger = \tau_\beta \tau_{\beta_{\rm in}} \tau_\beta.
        \end{equation}
        Therefore, the overall output of the quantum SWITCH is given by:
        \begin{equation}
            \big[\mathcal{S}(\mathcal{T}_{\beta}, ..., \mathcal{T}_{\beta})\big] (\tau_{\beta_{\rm in}} \otimes \omega) = \sum_i \tau_\beta \otimes \omega_{ii} |i\rangle\langle i| + \sum_{i\neq j} \tau_\beta \tau_{\beta_{\rm in}} \tau_\beta \otimes \omega_{ij} |i\rangle \langle j|.
        \end{equation}
        Performing some algebraic manipulations, we can rewrite it as:
        \begin{equation}
            \big[\mathcal{S}(\mathcal{T}_{\beta}, ..., \mathcal{T}_{\beta})\big] (\tau_{\beta_{\rm in}} \otimes \omega) = \frac{1}{2} \Bigl( (\tau_\beta + \tau_\beta \tau_{\beta_{\rm in}} \tau_\beta) \otimes \omega + (\tau_\beta - \tau_\beta \tau_{\beta_{\rm in}} \tau_\beta) \otimes \tilde{\omega} \Bigr),
        \end{equation}
        where $\tilde{\omega}$ is a state that is obtained by inverting the sign of off-diagonal elements of $\omega$. Denoting the unnormalized states of work medium by $\sigma_\pm$, we obtain~(\ref{app:eq:NOutput}). hence the proof.
    \end{proof}
\end{lemma}

Finally, in the following Proposition, we provide the optimal initial state and calculate the maximal SWITCH-ergotropy.

\begin{proposition}
     The SWITCH-ergotropy~(\ref{def:SwitchErg}) with uncorrelated joint input state is maximal for
     \begin{equation}\label{app:eq:NUCoptstate}
          \rho_{SC}^{\rm opt} = \tau_{\beta_{\rm in}} \otimes \ket{\gamma_+}\bra{\gamma_+},
    \end{equation}
    where $|\gamma_+\rangle = \frac{1}{\sqrt{N}} \sum_{i=0}^{N-1} |i\rangle$, and is given by:
    \begin{equation}\label{app:eq:NErgUC}
    \mathcal{W}_{\mathrm{UC}}^{N} = \frac{N-1}{NZ_{S,\beta}^2Z_{S,\beta_{\rm in}}} \operatorname{max}\{ 0, e^{-2\beta} - e^{-\beta_{\rm in}}\}.
    \end{equation}
\begin{proof}    
Applying Lemma~\ref{app:lemma:NUCoutput}, we calculate the SWITCH-ergotropy due to a two-outcome measurement $\{ \Pi, \mathds{1} - \Pi \}$ with respect to a projector $\Pi$:
\begin{align}
    \nonumber \mathcal{W}_{\rm UC}^{N} &= \max_{\Pi, \omega} \Biggl( W\Bigl( \operatorname{Tr}[\Pi\omega] \sigma_+ + \operatorname{Tr}[\Pi\tilde{\omega}] \sigma_- \Bigr) \\
    &+ W\Bigl( (1 - \operatorname{Tr}[\Pi\omega]) \sigma_+ + (1 - \operatorname{Tr}[\Pi\tilde{\omega}]) \sigma_- \Bigr) \Biggr).
\end{align}
Taking into account that $\omega + \tilde{\omega} = 2\operatorname{diag}[\omega]$ and $\omega - \tilde{\omega} = 2\operatorname{off-diag}[\omega]$, where $\operatorname{diag}[\omega]$ is the matrix with the only non-zero entries being diagonal elements of $\omega$ and $\operatorname{off-diag}[\omega]$ is the matrix with the only non-zero entries being off-diagonal ones, we can expand the SWITCH-ergotropy as
\begin{align}
    \mathcal{W}_{\rm UC}^{N} &= \frac{1}{2} \max_{\Pi, \omega} \Biggl( W\Bigl( \tau_\beta + G(\Pi, \omega) \Bigr) + W\Bigl( \tau_\beta - G(\Pi, \omega) \Bigr) \Biggr), \\
    G(\Pi, \omega) &= \Bigl( 1 - 2\operatorname{Tr}\Bigl[\Pi \operatorname{diag}[\omega] \Bigr]\Bigr) \tau_\beta - 2\operatorname{Tr}\Bigl[\Pi \operatorname{off-diag}[\omega]\Bigr] \tau_\beta \tau_{\beta_{\rm in}} \tau_\beta.
\end{align}
Since the term $G(\Pi, \omega)$ does not carry quantum coherence in the energetic basis, the corresponding ergotropy consists of \textit{incoherent} counterpart only, which can be calculated by applying~(\ref{app:DaemIncoherErg}):
\begin{align}
    \nonumber \mathcal{W}_{\rm UC}^{N} &= \frac{1}{2Z_{S,\beta}}\max_{\Pi, \omega} \operatorname{max}\Biggl\{ 0, \Biggl| \Bigl( 1 - 2\operatorname{Tr}\Bigl[\Pi \operatorname{diag}[\omega] \Bigr]\Bigr)( 1 - e^{-\beta}) \\
    &- 2\operatorname{Tr}\Bigl[\Pi \operatorname{off-diag}[\omega]\Bigr] \frac{1 - e^{-(2\beta + \beta_{\rm in})}}{Z_{S,\beta} Z_{S,\beta_{\rm in}}} \Biggr| - (1 - e^{-\beta}) \Biggr\}.
\end{align}
Taking into account that $\operatorname{diag}[\omega] + \operatorname{off-diag}[\omega] = \omega$, we can rewrite the SWITCH-ergotropy as:
\begin{eqnarray}
    \nonumber \mathcal{W}_{\rm UC}^{N} &=& \frac{1}{2Z_{S,\beta}^2 Z_{S,\beta_{\rm in}}}\max_{\Pi, \omega} \operatorname{max}\Biggl\{ 0, \Bigl| \Bigl( 2\operatorname{Tr}\Bigl[\Pi \operatorname{diag}[\omega] \Bigr] - 1 \Bigr)( e^{-2\beta} - e^{-\beta_{\rm in}}) \\
    &-& \Bigl( 2\operatorname{Tr}[\Pi \omega ] - 1 \Bigr) (1 - e^{-(2\beta + \beta_{\rm in})}) \Bigr| - (1 - e^{-2\beta})(1 + e^{-\beta_{\rm in}}) \Biggr\},
\end{eqnarray}
This expression for the SWITCH-ergotropy allows us to separate two cases with respect to the temperature bound~(\ref{eq:TempLimit}). If $\beta_{\rm in} \leq 2\beta$, we obtain:
\begin{align}\label{app:eq:NErgNonopt_zero}
    \nonumber \mathcal{W}_{\rm UC}^{N, \beta_{\rm in} \leq 2\beta} &= \frac{1}{Z_{S,\beta}^2 Z_{S,\beta_{\rm in}}} \max_{\Pi, \omega} \max\Biggl\{0, -\Bigl( 1 - \operatorname{Tr}\Bigl[\Pi \operatorname{diag}[\omega] \Bigr]\Bigr) ( e^{-\beta_{\rm in}} - e^{-2\beta}) \\
    &- \Bigl( 1 - \operatorname{Tr}[\Pi \omega ]\Bigr) (1 - e^{-(2\beta + \beta_{\rm in})}) \Biggr\}.
\end{align}    
Since all the factors~(\ref{app:eq:NErgNonopt_zero}) are non-negative, the presence of minus signs makes the overall SWITCH-ergotropy zero regardless of $\omega$ and $\Pi$. On the other hand, for $\beta_{\rm in} \geq 2\beta$, we obtain:
\begin{align}\label{app:eq:NErgNonopt}
    \nonumber \mathcal{W}_{\rm UC}^{N, \beta_{\rm in} > 2\beta} &= \frac{1}{Z_{S,\beta}^2 Z_{S,\beta_{\rm in}}} \max_{\Pi, \omega} \Biggl( \Bigl( 1 - \operatorname{Tr}\Bigl[\Pi \operatorname{diag}[\omega] \Bigr]\Bigr) ( e^{-2\beta} - e^{-\beta_{\rm in}}) \\
    &- \Bigl( 1 - \operatorname{Tr}[\Pi \omega ]\Bigr) (1 - e^{-(2\beta + \beta_{\rm in})}) \Biggr).
\end{align}
As all the factors in~(\ref{app:eq:NErgNonopt}) are non-negative, its optimization requires maximization of $\operatorname{Tr}[\Pi \omega ]$ and minimization of $\operatorname{Tr}[\Pi \operatorname{diag}[\omega] ]$. The first condition means that $\omega$ is a pure state, and the two-outcome measurement of the control system is performed with respect to the projector $\Pi = \omega \equiv |\psi\rangle\langle\psi|$. In turn, minimization of $\operatorname{Tr}[\Pi\operatorname{diag}[\omega]] \equiv \langle\psi | \operatorname{diag}[|\psi \rangle\langle\psi |]|\psi\rangle$ is equivalent to minimization of $\sum_j |a_j|^4$, where $a_j$ are amplitudes of $|\psi\rangle = \sum_j a_j |j\rangle$. It is achieved if all amplitudes are equal up to phases, i.e., $|\psi\rangle = \frac{1}{\sqrt{N}} \sum_{j=0}^{N-1} e^{i\phi_j} |j\rangle$. Therefore, without loss of generality, we can choose the initial state $\rho_{SC} = \tau_{\beta_{\rm in}} \otimes |\gamma_+\rangle\langle\gamma_+|$, where $|\gamma_+\rangle = \frac{1}{\sqrt{N}} \sum_{j=0}^{N-1} |j\rangle$, which optimizes the SWITCH-ergotropy:
\begin{equation}
    \mathcal{W}_{\rm UC}^{N} = \frac{1}{Z_{S,\beta}^2 Z_{S,\beta_{\rm in}}} \Bigl( 1 - \frac{1}{N}\Bigr) \max\{ 0,  e^{-2\beta} - e^{-\beta_{\rm in}} \}.
\end{equation}
This leads to~(\ref{app:eq:NErgUC}), hence the proof.
\end{proof}
\end{proposition}

\section{Proof of Theorem~\ref{theor:ErgCC}}
\label{appCC}

In this Appendix, we find the maximal SWITCH-ergotropy that can be achieved under initial classical correlations between work medium in a locally thermal state and the control system. We start by the following two Lemmas that will simplify the proof of Theorem~\ref{theor:ErgCC}.

\begin{lemma}\label{lemma:app:CCIdentChan}
    Given a classically correlated joint initial state $\rho_{SC} = \sum_{ij} p_{ij} \Pi_i \otimes \Theta_j$ and $2$ identical channels $\mathcal{E}[\cdot]$, the action of the quantum SWITCH controlling their causal order can be written as:
    \begin{align}\label{app:CCswitch}
        \nonumber \big[ \map S (\map{E}, \map{E}) \big]  (\rho_{SC}) &=  \frac{1}{4} \sum_{ij} p_{ij} \sum_{i'j'} \Bigl( \{E_{i'}, E_{j'}\} \Pi_i \{E_{i'}, E_{j'}\}^\dagger \otimes \Theta_j \\
        &+ [E_{i'}, E_{j'}] \Pi_i [E_{i'}, E_{j'}]^\dagger \otimes Z\Theta_j Z \Bigr).
    \end{align}
    where $\{E_i\}_i$ is the Kraus decomposition of $\mathcal{E}$.
    \begin{proof}
        The proof follows immediately by observation that quantum SWITCH produces a CPTP map that acts linearly on $\rho_{SC}$, which is a convex combination of states $\Pi_i \otimes \Theta_j$, and application of Lemma~\ref{lemma:app:IdentChan} to each of them.
    \end{proof}
\end{lemma}

\begin{lemma}
    Given a qubit work medium $S$, its initial locally thermal state with respect to condition~(\ref{eq:LocalThermality}) can be given as:
    \begin{align}\label{app:rhoThermalCC}
        \nonumber \rho_{SC} &= \ket{\chi}\bra{\chi}\otimes \Bigl[ p_0 \ket{\phi}\bra{\phi} + \Bigl( \frac{1}{Z_{S,\beta_{\rm in}}} - p_0 \Bigr) \ket{\phi^\perp}\bra{\phi^\perp} \Bigr] \\
        &+ \ket{\chi^\perp}\bra{\chi^\perp}\otimes \Bigl[ p_1 \ket{\phi}\bra{\phi} + \Bigl( \frac{e^{-\beta_{\rm in}} }{Z_{S,\beta_{\rm in}}} - p_1 \Bigr) \ket{\phi^\perp}\bra{\phi^\perp} \Bigr] ,
    \end{align}
    where $\{ \ket{\chi}, \ket{\chi^\perp}\}$ constitute an orthonormal basis for the work medium, which is given by $\{ \ket{0}, \ket{1}\}$ for $\beta_{\rm in} > 0$ and arbitrary for $\beta_{\rm in} = 0$, $\{ \ket{\phi}, \ket{\phi^\perp}\}$ constitute an arbitrary orthonormal basis for the control qubit, and $0 \leq p_0 \leq \frac{1}{Z_{S,\beta_{\rm in}}}$ and $0 \leq p_1 \leq \frac{e^{-\beta_{\rm in}}}{Z_{S,\beta_{\rm in}}}$.
\begin{proof}
    By definition, a state of the joint system $SC$ that reveals completely classical correlations, can be given as
    \begin{equation}\label{app:rhoCC}
        \rho_{SC} = \sum_{ij} p_{ij} \Pi_i \otimes \Theta_j,
    \end{equation}
    where $\{\Pi_i\}$ and $\{\Theta_j\}$ are rank-1 projectors with respect to an orthonormal basis of $S$ and $C$, respectively, and $\sum_{ij} p_{ij} = 1$. Taking into account the condition~(\ref{eq:LocalThermality}) of local thermality of $S$, we obtain
    \begin{equation}\label{eq:AppCCExpansion}
        \operatorname{Tr}_C[\rho_{SC}] = \sum_i \Bigl( \sum_j p_{ij} \Bigr) \Pi_i \equiv \tau_{\beta_{\rm in}},
    \end{equation}
    Writing down explicitly
    \begin{equation}\label{app:thermstate}
        \tau_{\beta_{\rm in}} = \frac{1}{Z_{\beta_{\rm in}}} \Bigl( |0\rangle\langle 0| + e^{-\beta_{\rm in}} |1\rangle\langle 1| \Bigr),
    \end{equation}
    we find that the set of projectors $\{ \Pi_i \}_i = \{ |\chi\rangle\langle \chi|, |\chi^\perp\rangle\langle \chi^\perp| \}$ is fixed as $\{ |0\rangle\langle 0|, |1 \rangle\langle 1 | \}$ as far as $\beta_{\rm in} > 0$. On the other hand, for $\beta_{\rm in} = 0$, the thermal state $\tau_{\beta_{\rm in}}$ is a maximally mixed state, hence, invariant under transformations of basis. Therefore, in this case, the set $\{ |\chi\rangle\langle \chi|, |\chi^\perp\rangle\langle \chi^\perp| \}$ is arbitrary.
    
    Plugging in~(\ref{app:thermstate}) into~(\ref{eq:AppCCExpansion}), we obtain the following conditions on $p_{ij}$,
    \begin{eqnarray}
        \sum_j p_{0j} = \frac{1}{Z_{\beta_{\rm in}}}, \\
        \sum_j p_{1j} = \frac{e^{-\beta_{\rm in}}}{Z_{\beta_{\rm in}}}.
    \end{eqnarray}
    By denoting $p_{00} \equiv p_0$ and $p_{10} \equiv p_1$, the joint initial state~(\ref{app:rhoThermalCC}) follows, hence the proof.
\end{proof}
\end{lemma}

Since the case $\beta_{\rm in} = 0$ features a degeneracy in decomposition of the joint classically correlated state and, hence, an optimization over all such decompositions is necessary, we analyze this case separately and focus firstly on the case $\beta_{\rm in} > 0$. In the following Proposition, we find the initial state that provides the maximal SWITCH-ergotropy.

\begin{proposition}\label{app:prop:OptStateCC}
     For $\beta_{\rm in} > 0$, the SWITCH-ergotropy~(\ref{def:SwitchErg}) with classically correlated joint input state is maximal for
     \begin{equation}
          \rho_{SC}^{\rm opt} = \frac{1}{Z_{S,\beta_{\rm in}}} \Bigl( \ket{0}\bra{0} \otimes \ket{+}\bra{+} + e^{-\beta_{\rm in}} \ket{1}\bra{1}\otimes \ket{-}\bra{-}\Bigr).
    \end{equation}
    
\begin{proof}

We start by plugging in the Kraus decomposition~(\ref{eq:ThermKraus}) of identical thermalizing channels and the initial joint state~(\ref{app:rhoThermalCC}) into~(\ref{app:CCswitch}), we obtain
\begin{align}
    \nonumber \big[ \map S (\mathcal{T}_\beta, \mathcal{T}_\beta) \big]  (\rho_{SC}) &=  \frac{1}{2}\Bigl\{ ( \tau_\beta + \tau_\beta |0\rangle\langle 0| \tau_\beta ) \otimes \Bigl[ p_0 \ket{\phi}\bra{\phi} + \Bigl( \frac{1}{Z_{S,\beta_{\rm in}}} - p_0 \Bigr) \ket{\phi^\perp}\bra{\phi^\perp} \Bigr] \\
    \nonumber &+ ( \tau_\beta + \tau_\beta |1\rangle\langle 1| \tau_\beta ) \otimes \Bigl[ p_1 \ket{\phi}\bra{\phi} + \Bigl( \frac{e^{-\beta_{\rm in}} }{Z_{S,\beta_{\rm in}}} - p_1 \Bigr) \ket{\phi^\perp}\bra{\phi^\perp} \Bigr] \\
   \nonumber  &+ ( \tau_\beta - \tau_\beta |0\rangle\langle 0| \tau_\beta ) \otimes \Bigl[ p_0 Z\ket{\phi}\bra{\phi}Z + \Bigl( \frac{1}{Z_{S,\beta_{\rm in}}} - p_0 \Bigr) Z\ket{\phi^\perp}\bra{\phi^\perp}Z \Bigr] \\
    \nonumber &+ ( \tau_\beta - \tau_\beta |1\rangle\langle 1| \tau_\beta ) \otimes \Bigl[ p_1 Z\ket{\phi}\bra{\phi}Z + \Bigl( \frac{e^{-\beta_{\rm in}} }{Z_{S,\beta_{\rm in}}} - p_1 \Bigr) Z\ket{\phi^\perp}\bra{\phi^\perp}Z \Bigr] \Bigr\}.
\end{align}
Taking into account the identities $\ket{\phi/\phi^\perp}\bra{\phi/\phi^\perp} + Z\ket{\phi/\phi^\perp}\bra{\phi/\phi^\perp}Z = \mathds{1} \pm \langle Z \rangle_\phi Z $ and $\ket{\phi/\phi^\perp}\bra{\phi/\phi^\perp} - Z\ket{\phi/\phi^\perp}\bra{\phi/\phi^\perp}Z = \pm \Bigl(\langle X \rangle_\phi X + \langle Y \rangle_\phi Y \Bigr)$, it can be simplified as:
\begin{equation}\label{app:CCThermalswitch}
    \big[ \map S (\mathcal{T}_\beta, \mathcal{T}_\beta) \big]  (\rho_{SC}) =  \frac{1}{2}\Bigl\{  \tau_\beta \otimes \Bigl( \mathds{1} + (2p_0 + 2p_1 - 1) \langle Z \rangle_\phi Z \Bigr) + G(p_0, p_1) \otimes \Bigl( \langle X \rangle_\phi X + \langle Y \rangle_\phi Y \Bigr) \Bigr\},
\end{equation}
where $G(p_0, p_1) = \tau_\beta \Bigl[ \Bigl(2p_0 - \frac{1}{Z_{S,\beta_{\rm in}}} \Bigr) |0\rangle\langle 0| + \Bigl(2p_1 - \frac{e^{-\beta_{\rm in}}}{Z_{S,\beta_{\rm in}}} \Bigr) |1\rangle\langle 1| \Bigr] \tau_\beta$. The corresponding maximal SWITCH-ergotropy can be given as:
\begin{eqnarray}\label{app:eq:CCerg}
    \nonumber \mathcal{W}_{\rm CC}^{\beta_{\rm in} > 0} &=& \frac{1}{2}\max_{p_0, p_1, |\phi\rangle, |\psi\rangle} \Biggl( W\Bigl( (1+\tilde{p}(p_0, p_1, \psi, \omega)) \tau_\beta + \lambda (\psi, \omega) G(p_0, p_1) \Bigr) \\
    &+& W\Bigl( (1 - \tilde{p}(p_0, p_1, \psi, \omega)) \tau_\beta - \lambda(\psi, \omega) G(p_0, p_1) \Bigr)\Biggr),
\end{eqnarray}
where $\tilde{p}(p_0, p_1, \psi, \omega) = (2p_0 + 2p_1 - 1) \langle Z \rangle_\phi \langle Z \rangle_\psi$ and $\lambda(\phi, \psi) = \langle X \rangle_\phi \langle X \rangle_\psi + \langle Y \rangle_\phi \langle Y \rangle_\psi$. As $\tau_\beta$ is a thermal state, hence, carrying zero ergotropy, the only possible source of non-zero ergotropy in~(\ref{app:eq:CCerg}) are the terms proportional to $G(p_0, p_1)$. Therefore, optimization of the SWITCH-ergotropy~(\ref{app:eq:CCerg}) is achieved by maximal value of $\lambda(\psi, \omega)$, i.e., by the states $|\phi\rangle$ and $|\psi\rangle$ with $|\langle X \rangle_\phi| = |\langle X \rangle_\psi| = 1$ or $|\langle Y \rangle_\phi| = |\langle Y \rangle_\psi| = 1$. This means that the optimal set of projectors $\{\Theta_j\}_j$ in~(\ref{app:rhoCC}) can be taken as one composed by the states $|\pm\rangle = \frac{|0\rangle \pm |1\rangle}{\sqrt{2}}$ or $|\pm i\rangle = \frac{|0\rangle \pm i |1\rangle}{\sqrt{2}}$. Without loss of generality, we stick to $\{\Theta_j\}_j = \{ |+\rangle\langle +|, |-\rangle\langle -|\}$.

In order to find optimal $p_0$ and $p_1$, we take into account that $G$ does not carry coherence. Therefore, the resulting ergotropy (if any) consists only of its incoherent counterpart, which is optimized by maximizing the difference $|\delta G|$ of energetic populations of $G$. It is easy to see that it is achieved by taking maximal $p_0$ and minimal $p_1$ or vice versa, i.e., $p_0 = \frac{1}{Z_{S,\beta_{\rm in}}}, p_1 = 0$ or $p_0 = 0, p_1 = \frac{e^{-\beta_{\rm in}}}{Z_{S,\beta_{\rm in}}}$. Therefore, the optimal classically correlated state reads
\begin{equation}\label{app:rhoThermalCCOpt}
\rho_{SC}^{\rm opt} = \frac{1}{Z_{S,\beta_{\rm in}}} \Bigl( \ket{0}\bra{0} \otimes \ket{+}\bra{+} + e^{-\beta_{\rm in}} \ket{1}\bra{1}\otimes \ket{-}\bra{-}\Bigr),
\end{equation}
hence the proof.
\end{proof}
\end{proposition}

Application of Proposition~\ref{app:prop:OptStateCC} allows us to calculate the resulting maximal SWITCH-ergotropy, as shown in the following Proposition.
\begin{proposition}
For $\beta_{\rm in} > 0$, the maximal SWITCH-ergotropy under classical correlations is given by:
\begin{equation}\label{app:prop:ErgCC}
        \mathcal{W}_{\mathrm{CC}}^{\beta_{\rm in} > 0} = \frac{\operatorname{max}\{ 0, e^{-2\beta} - e^{-\beta_{\rm in}} + 2e^{-(2\beta + \beta_{\rm in})}\}}{2Z_{S,\beta}^2Z_{S,\beta_{\rm in}}},
    \end{equation}
\begin{proof}
Applying Proposition~\ref{app:prop:OptStateCC}, we take $|\psi\rangle = |\phi\rangle = |+\rangle$, so that $\langle X \rangle_\psi = \langle X \rangle_\phi = 1$ and $\langle Y \rangle_\psi = \langle Y \rangle_\phi = \langle Z \rangle_\psi = \langle Z \rangle_\phi = 0$, and $p_0 = \frac{1}{Z_{S, \beta_{\rm in}}}$ and $p_1 = 0$ in~(\ref{app:eq:CCerg}), obtaining hence the maximal SWITCH-ergotropy under classical correlations
\begin{equation}
    \mathcal{W}_{\rm CC}^{\beta_{\rm in} > 0} = \frac{1}{2} W( \tau_\beta + G ) + \frac{1}{2} W( \tau_\beta - G ),
\end{equation}
where $G = \tau_\beta \Bigl[ \frac{1}{Z_{S,\beta_{\rm in}}} |0\rangle\langle 0| - \frac{e^{-\beta_{\rm in}}}{Z_{S,\beta_{\rm in}}}|1\rangle\langle 1| \Bigr] \tau_\beta$. Since $G$ does not carry quantum coherence in the energetic basis, the corresponding ergotropy consists of \textit{incoherent} counterpart only, which can be given as~\cite{Simonov2022}
\begin{equation}
    \mathcal{W}_{\rm CC}^{\beta_{\rm in} > 0} = \frac{1}{2}\{0, |\delta G| - |\delta \rho_{\rm def}|\},
\end{equation}
where $\delta G$ is the difference of the energetic populations of $G$, and $\delta \rho_{\rm def}$ is the difference of the energetic populations of $\rho_{\rm def} = \tau_\beta$ of the output of causally separable combination of $\mathcal{T}_\beta$, i.e., $(\mathcal{T}_\beta \circ \mathcal{T}_\beta)$. Therefore, quantum SWITCH activates ergotropy
\begin{eqnarray}
     \mathcal{W}_{\rm CC}^{\beta_{\rm in} > 0} &=& \frac{1}{2Z_{S,\beta}^2 Z_{S,\beta_{\rm in}}}\operatorname{max}\Bigl\{ 0, 1 + e^{-(2\beta + \beta_{\rm in})} - Z_{S,\beta} Z_{S,\beta_{\rm in}} (1 - e^{-\beta})\Bigr\} \\
     &=& \frac{1}{2Z_{S,\beta}^2 Z_{S,\beta_{\rm in}}}\operatorname{max}\Bigl\{ 0, e^{-2\beta} - e^{-\beta_{\rm in}} + 2e^{-(2\beta + \beta_{\rm in})} \Bigr\},
\end{eqnarray}
hence the proof. 
\end{proof}
\end{proposition}

Finally, in the following Proposition, we provide the optimal initial state and calculate the maximal SWITCH-ergotropy for the case $\beta_{\rm in} = 0$.

\begin{proposition}\label{app:prop:OptStateCCZero}
     For $\beta_{\rm in} = 0$, the SWITCH-ergotropy~(\ref{def:SwitchErg}) with classically correlated joint input state is maximal for
     \begin{equation}\label{app:eq:CCoptstate:zero}
          \rho_{SC}^{\rm opt} = \frac{1}{2} \Bigl( \ket{+}\bra{+} \otimes \ket{+}\bra{+} + \ket{-}\bra{-}\otimes \ket{-}\bra{-}\Bigr),
    \end{equation}
    and is given by:
    \begin{equation}\label{app:eq:ErgCC:zero}
    \mathcal{W}_{\mathrm{CC}}^{\beta_{\rm in} = 0} = \frac{1}{2}\tanh\Bigl(\frac{\beta}{2}\Bigr)\Bigl( \sqrt{1 + \frac{1}{4}\sinh^{-2}(\beta)} - 1\Bigr).
    \end{equation}

\begin{proof}
Following the same steps as in the proof of Proposition~\ref{app:prop:OptStateCC}, we can write the output of the quantum SWITCH as:
\begin{equation}\label{app:CCThermalswitchZero}
    \nonumber \big[ \map S (\mathcal{T}_\beta, \mathcal{T}_\beta) \big]  (\rho_{SC}) =  \frac{1}{2}\Bigl\{  \tau_\beta \otimes \Bigl( \mathds{1} + (2p_0 + 2p_1 - 1) \langle Z \rangle_\phi Z \Bigr) + G(p_0, p_1) \otimes \Bigl( \langle X \rangle_\phi X + \langle Y \rangle_\phi Y \Bigr) \Bigr\},
\end{equation}
where $G(p_0, p_1) = \tau_\beta \Bigl[ \Bigl(2p_0 - \frac{1}{2} \Bigr) |\chi\rangle\langle \chi| + \Bigl(2p_1 - \frac{1}{2} \Bigr) |\chi^\perp\rangle\langle \chi^\perp| \Bigr] \tau_\beta$. Since $G(p_0, p_1)$ is the only term that can contribute to ergotropy, similarly to Proposition~\ref{app:prop:OptStateCC}, we can take $|\psi\rangle = |\phi\rangle = |+\rangle$ in order to maximize it. However, in contrast to Proposition~\ref{app:prop:OptStateCC}, the resulting ergotropy can have coherent counterpart as $G(p_0, p_1)$ can carry coherence as far as $|\chi\rangle \neq |0\rangle$. Hence, we can write it as:
\begin{equation}\label{app:eq:CCergBetaZero}
    \mathcal{W}_{\rm CC}^{\beta_{\rm in} = 0} = \frac{1}{2}\max_{p_0, p_1} \Biggl( W\Bigl( \tau_\beta + G(p_0, p_1) \Bigr) + W\Bigl( \tau_\beta - G(p_0, p_1) \Bigr)\Biggr).
\end{equation}
Both incoherent and coherent counterparts are maximized by taking maximal $p_0$ and minimal $p_1$ or vice versa, i.e. $p_0 = \frac{1}{2}$ and $p_1 = 0$ or $p_0 = 0$ and $p_1 = \frac{1}{2}$. Without loss of generality, we stick to the former and, expanding $|\chi\rangle = \cos\frac{\theta}{2} |0\rangle + e^{i\phi}\sin\frac{\theta}{2} |1\rangle$, calculate separately the incoherent and coherent counterparts of~(\ref{app:eq:CCergBetaZero}) by applying~(\ref{eq:IncohErg}) and~(\ref{eq:CohErg}), respectively:
\begin{align}\label{app:CCoptimized}
    \mathcal{W}_{\rm CC}^{\beta_{\rm in} = 0} &= \max_\theta \Bigl( W_{\rm inc}^{\rm CC}(\theta) + W_{\rm coh}^{\rm CC}(\theta) \Bigr), \\
    \label{app:CCoptimized:incoh} W_{\rm inc}^{\rm CC}(\theta) &= \frac{1}{4Z_{S,\beta}^2}\max\Bigl\{0, |\cos(\theta)|(1 + e^{-2\beta}) - 2(1 - e^{-2\beta})\Bigr\}, \\
    \label{app:CCoptimized:coh} W_{\rm coh}^{\rm CC}(\theta) &= \frac{\tanh\Bigl(\frac{\beta}{2}\Bigr)}{4}\Biggl(\sqrt{\zeta_+^2(\theta) + \frac{\sin^2(\theta)}{4\sinh^2(\beta)}} - | \zeta_+(\theta)| + \sqrt{\zeta_-^2 (\theta) + \frac{\sin^2(\theta)}{4\sinh^2(\beta)}} - | \zeta_- (\theta) | \Biggr),
\end{align}
where $\zeta_\pm (\theta) = 1 \pm \cos\theta\frac{\coth\beta}{2}$. Optimization of ergotropy~(\ref{app:CCoptimized}) shows that the optimal angle is given by $\theta = \frac{\pi k}{2}$, $k \in \mathbb{Z}$, i.e., it is maximized by states $|\chi\rangle = \frac{1}{\sqrt{2}}(|0\rangle + e^{i\phi}|1\rangle)$. Without loss of generality, we fix $\phi = 0$, hence, obtaining $|\chi\rangle = |+\rangle$ and the optimal joint state~(\ref{app:eq:CCoptstate:zero}). In turn, the incoherent counterpart~(\ref{app:CCoptimized:incoh}) of ergotropy becomes zero, hence, only the coherent counterpart~(\ref{app:CCoptimized}) contributes to it. A straightforward calculation leads to~(\ref{app:eq:ErgCC:zero}), hence the proof.

\end{proof}
\end{proposition}

\section{Proof of Theorem~\ref{theor:ErgDiscord}}
\label{appDiscord}

In this Appendix, we find the maximal SWITCH-ergotropy that can be achieved under separable initial joint state of work medium (under the condition of local thermality) and the control system, 
\begin{eqnarray}\label{app:eq:SepState}
    \rho_{SC} &=& \sum_i p_i \rho_i \otimes \omega_i, \\
    \sum_i p_i \rho_i &=& \tau_{\beta_{\rm in}},
\end{eqnarray}
so that it can carry non-zero quantum discord. We start by the following three Lemmas that will simplify the proof of Theorem~\ref{theor:ErgDiscord}.

\begin{lemma}\label{lemma:app:DIdentChan}
    Given a separable joint initial state~(\ref{app:eq:SepState}) and $2$ identical channels $\mathcal{E}[\cdot]$, the action of the quantum SWITCH controlling their causal order can be written as:
    \begin{equation}\label{app:Dswitch}
        \big[ \map S (\map{E}, \map{E}) \big]  (\rho_{SC})  =  \frac{1}{4} \sum_i p_i \sum_{i'j'} \Bigl( \{E_{i'}, E_{j'}\} \rho_i \{E_{i'}, E_{j'}\}^\dagger \otimes \omega_i + [E_{i'}, E_{j'}] \rho_i [E_{i'}, E_{j'}]^\dagger \otimes Z\omega_i Z \Bigr).
    \end{equation}
    where $\{E_i\}_i$ is the Kraus decomposition of $\mathcal{E}$.
    \begin{proof}
        The proof follows immediately by observation that quantum SWITCH produces a CPTP map that acts linearly on $\rho_{SC}$, which is a convex combination of states $\rho_i \otimes \omega_i$, and application of Lemma~\ref{lemma:app:IdentChan} to each of them.
    \end{proof}
\end{lemma}

Now, in the following Lemma, we show that it is enough to consider only pure states $\rho_i$ in order to find the maximal SWITCH-ergotropy. 

\begin{lemma}\label{lemma:app:DPureStates}
    The separable joint state $\rho_{SC}$ that maximizes the SWITCH-ergotropy~(\ref{def:SwitchErg}) has the form:
    \begin{equation}\label{app:eq:PureDiscord}
        \rho_{SC}^{\rm opt} = p |\mu \rangle \langle \mu | \otimes |+ \rangle \langle + | + (1-p) |\nu \rangle \langle \nu | \otimes |- \rangle \langle - |,
    \end{equation}
    where $|\mu/\nu\rangle$ are certain not necessarily orthogonal pure states satisfying the condition of local thermality $p |\mu \rangle \langle \mu | + (1-p) |\nu \rangle \langle \nu | = \tau_{\beta \rm in}$.
    \begin{proof}
        First, we find the output of the quantum SWITCH by applying Lemma~\ref{lemma:app:IdentChan}:
        \begin{equation}\label{eq:app:IdentChanD}
            \big[ \map S (\mathcal{T}_\beta, \mathcal{T}_\beta) \big]  (\rho_{SC}) = \sum_i p_i \Bigl[ \sigma_{+, i} \otimes \omega_i + \sigma_{-, i} \otimes Z\omega_i Z \Bigr],
        \end{equation}
        where 
        \begin{align}
            \nonumber \sigma_{+, i} &= \frac{1}{4} \sum_{i'j'i''j''} \{T_{\beta, i'j'}, T_{\beta, i''j''}\} \rho_i \{T_{\beta, i'j'}, T_{\beta, i''j''}\}^\dagger, \\
            \nonumber \sigma_{-, i} &= \frac{1}{4} \sum_{i'j'i''j''} [T_{\beta, i'j'}, T_{\beta, i''j''}] \rho_i [T_{\beta, i'j'}, T_{\beta, i''j''}]^\dagger.
        \end{align}
        The maximal SWITCH-ergotropy is given then by optimization of its daemonic ergotropy over all sets of states $\omega_i$. Applying the definition of daemonic ergotropy~(\ref{DaemonicErg}), we can write the maximal SWITCH-ergotopy as:
        \begin{align}
            \nonumber \mathcal{W}_{\rm D} &= \max_{\rho_i, \omega_i, |\psi\rangle} \Biggl(  W\Bigl(\sum_i p_i \Bigl[ \langle \psi | \omega_i | \psi \rangle \sigma_{+, i} + \langle \psi | Z\omega_i Z | \psi \rangle \sigma_{-, i} \Bigr]\Bigr) \\
            &+ W\Bigl(\sum_i p_i \Bigl[\langle \psi^\perp | \omega_i | \psi^\perp \rangle \sigma_{+, i} + \langle \psi^\perp | Z\omega_i Z | \psi^\perp \rangle \sigma_{-, i} \Bigr] \Bigr)\Biggr),
        \end{align}
        where $\{|\psi\rangle, |\psi^\perp\rangle\}$ form the measurement basis. On the other hand, we can rewrite $\sigma_{+, i} + \sigma_{-, i} = \tau_\beta$, hence, obtaining
        \begin{align}
            \nonumber \nonumber \mathcal{W}_\textrm{D} & = \frac{1}{2} \max_{\rho_i, \omega_i, |\psi\rangle} \Biggl( W\Bigl(\sum_i p_i [\langle \psi | (\omega_i + Z\omega_i Z) | \psi \rangle \tau_\beta + \langle \psi | (\omega_i - Z\omega_i Z) | \psi \rangle G_i ]\Bigr)\nonumber\\ &\quad + W\Bigl(\sum_i p_i [\langle \psi^\perp | (\omega_i + Z\omega_i Z) | \psi^\perp \rangle \tau_\beta + \langle \psi^\perp | (\omega_i - Z\omega_i Z) | \psi^\perp \rangle G_i ]\Bigr)\Biggr)
        \end{align}
        where $G_i \equiv \sigma_{+, i} - \sigma_{-, i}$. Taking into account that $\omega_i + Z\omega_i Z = \mathds{1} + \langle Z \rangle_{\omega_i} Z$ and $\omega_i - Z\omega_i Z = \langle X \rangle_{\omega_i} X + \langle Y \rangle_{\omega_i} Y$, where $\langle A \rangle_{\omega_i} = \operatorname{Tr}[A\omega_i]$, we find
        \begin{align}\label{app:eq:2ergD}
            \nonumber \mathcal{W}_{\rm D} &= \frac{1}{2}\max_{p_i, \rho_i, \omega_i, |\psi\rangle} \Biggl( W\Bigl( (1+\tilde{p}(\psi, \omega_i, p_i)) \tau_\beta + \sum_i p_i \lambda_i (\psi, \omega_i) G_i \Bigr) \\
            &+ W\Bigl( (1 - \tilde{p}(\psi, \omega_i, p_i)) \tau_\beta - \sum_i p_i \lambda_i(\psi, \omega_i) G_i \Bigr)\Biggr),
        \end{align}
        where $\tilde{p}(\psi, \omega_i, p_i) = \langle Z \rangle_\psi \sum_i p_i \langle Z \rangle_{\omega_i}$ and $\lambda(\psi, \omega_i) = \langle X \rangle_\psi \langle X \rangle_{\omega_i} + \langle Y \rangle_\psi \langle Y \rangle_{\omega_i}$. In turn, only terms proportional to $G_i$ can contribute to the SWITCH-ergotropy in~\ref{app:eq:2ergD}, therefore, its optimization is achieved by maximizing the absolute values of $\lambda_i(\psi, \omega_i)$, i.e., by the states $|\psi\rangle$ and $\omega_i$ with $|\langle X \rangle_\psi| = |\langle X \rangle_{\omega_i}| = 1$ or $|\langle Y \rangle_\psi| = |\langle Y \rangle_{\omega_i}| = 1$. This means that the optimal control states $\omega_i$ are pure and can be taken from the state sets $\{|\pm\rangle\}$ or $\{|\pm i\rangle\}$: without loss of generality, we stick to the first set, so that $\omega_i \in \{|\pm\rangle\}$. 

        Calculating $G_i$ as done in the previous Appendices, we obtain $G_i = \tau_\beta \rho_i \tau_\beta$. Therefore, applying the above optimization, we can rewrite the terms proportional to $G_i$ as:
        \begin{equation}
            \sum_i p_i \lambda_i(\psi, \omega_i) G_i = \tau_\beta \Bigl( \sum_i \operatorname{sign}(\lambda_i(\psi, \omega_i)) p_i \rho_i \Bigr) \tau_\beta.
        \end{equation}
        Two choices of signs of $\lambda_i(\psi, \omega_i)$ can be distinguished. First, due to the local thermality condition $\sum_i p_i \rho_i = \tau_{\beta_{\rm in}}$, the choice of the same sign of all $\lambda_i(\psi, \omega_i)$ corresponds to an initial product state featuring hence no a priori correlations and leading to the maximal SWITCH-ergotropy $\mathcal{W}_{\rm UC}^{N=2}$ guaranteed by Theorem~\ref{theor:ErgUC2}. On the other hand, the remaining option is to assign opposite signs to $\lambda_i(\psi, \omega_i)$. Indeed, denoting $I$ as the set of all indices $i$, we divide it into two disjoint sets $I_+ \sqcup I_- = I$, which correspond to the indices with $\lambda_i(\psi, \omega_i) = \pm 1$. Denoting $\sum_{i \in I_+} p_i \rho_i = p \rho_+$ and $\sum_{i \in I_-} p_i \rho_i = (1-p) \rho_-$ under local thermality condition $p \rho_+ + (1-p) \rho_- = \tau_{\beta_{\rm in}}$, we obtain an optimal initial state $\rho_{SC} = p \rho_+ \otimes |+\rangle \langle +| + (1-p) \rho_- \otimes |-\rangle \langle -|$ and the corresponding SWITCH-ergotropy:
        \begin{equation}
            \mathcal{W}_{\rm D} = \frac{1}{2}\max_{p, \rho_+} \Biggl( W\Bigl( \tau_\beta + G_{\rm D} \Bigr) + W\Bigl( \tau_\beta - G_{\rm D} \Bigr)\Biggr),
        \end{equation}
        where $G_{\rm D} = \tau_\beta \Bigr[ p  \rho_+ - (1-p) \rho_- \Bigr] \tau_\beta$. Applying the local thermality condition and writing $\rho_+ = \begin{pmatrix} p_+ & \rho_{01} \\ \rho_{01}^* & 1-p_+ \end{pmatrix}$, we can expand $G_{\rm D}$ as:
        \begin{equation}
            G_{\rm D} = \tau_\beta \Bigl[ \Bigl( 2p p_+ - \frac{1}{Z_{S, \beta_{\rm in}}} \Bigr) |0\rangle \langle 0| + \Bigl( 2p (1-p_+) - \frac{e^{-\beta_{\rm in}}}{Z_{S, \beta_{\rm in}}} \Bigr) |1\rangle \langle 1| + 2p (\rho_{01} |0\rangle \langle 1| + \rho_{01}^* |1\rangle \langle 0|) \Bigr] \tau_\beta.
        \end{equation}
        Comparing it with~(\ref{app:CCThermalswitch}) provided by initial classical correlations for $\beta_{\rm in} > 0$, we find that the first two terms coincide with $G(p_0, p_1)$ up to notation, i.e., $p_0 \equiv p p_+$ and $p_1 \equiv p (1 - p_+)$. Therefore, we find that the term $G_{\rm D}$ contributing exclusively to the SWITCH-ergotropy consists of an incoherent counterpart 
        \begin{equation}
        G_{\rm CC} \equiv \tau_\beta \Bigl[ \Bigl( 2p p_+ - \frac{1}{Z_{S, \beta_{\rm in}}} \Bigr) |0\rangle \langle 0| + \Bigl( 2p (1-p_+) - \frac{e^{-\beta_{\rm in}}}{Z_{S, \beta_{\rm in}}} \Bigr) |1\rangle \langle 1| \Bigr] \tau_\beta,
        \end{equation}
        coinciding with one of the maximal SWITCH-ergotropy under classical correlations guaranteed by Theorem~\ref{theor:ErgCC} and a coherent counterpart 
        \begin{equation}
        G_{\rm coh} \equiv \tau_\beta \Bigl[ 2p (\rho_{01} |0\rangle \langle 1| + \rho_{01}^* |1\rangle \langle 0|) \Bigr] \tau_\beta,
        \end{equation}
        due to the quantum coherence carried by $\rho_+$ and $\rho_-$. This means that maximal SWITCH-ergotropy under quantum discord is lower-bounded by the maximal SWITCH-ergotropy under classical correlations, which, in turn, is lower-bounded by the maximal SWITCH-ergotropy without a priori correlations. Therefore, the optimization of $\mathcal{W}_{\rm D}$ is achieved by the choice of opposite signs of $\lambda_i(\psi, \rho_i)$:
        \begin{equation}\label{app:eq:OptErgD}
            \mathcal{W}_{\rm D} = \frac{1}{2}\max_{p, \rho_+} \Biggl( W\Bigl( \tau_\beta + (G_{\rm CC} + G_{\rm coh}) \Bigr) + W\Bigl( \tau_\beta - (G_{\rm CC} + G_{\rm coh}) \Bigr)\Biggr).
        \end{equation}
        Finally, for fixed energetic populations of $\rho_+$ (i.e., its diagonal elements in energetic basis), quantum coherence contributes non-negatively to the SWITCH-ergotropy due to~(\ref{eq:CohErg}). Therefore, the maximal SWITCH-ergotropy is achieved by $\rho_+$ being a pure state. Therefore, it is enough to consider initial states $\rho_{SC}$ with pure states $|\mu/\nu\rangle$ correlated with the chosen states of the control qubit, i.e., $|\pm\rangle$, leading to the form~(\ref{app:eq:PureDiscord}) of the initial state, hence the proof.
    \end{proof}
\end{lemma}

In the following Lemma, we provide a simple form of the initial state $\rho_{SC}$ under local thermality condition that will help to calculate the maximal SWITCH-ergotropy. 

\begin{lemma}\label{app:lemma:locthermoptD}
    Under the local thermality condition, the separable joint state $\rho_{SC}$ that maximizes the SWITCH-ergotropy has the form:
    \begin{equation}\label{app:eq:locthermoptD}
        \rho_{SC}^{\rm opt} = \frac{1}{q + \mu - 2q\mu} \Bigl[ q(1-q) |\mu \rangle \langle \mu | \otimes |+\rangle\langle +| + \tau_{\beta_{\rm in}} |\mu^\perp \rangle \langle \mu^\perp | \tau_{\beta_{\rm in}} \otimes |-\rangle\langle -| \Bigr],
    \end{equation}
    where $\beta_{\rm in} = \ln\frac{q}{1-q}$.

    \begin{proof}    
Firstly, we apply Lemma~\ref{lemma:app:DPureStates} and write down explicitly:
\begin{eqnarray}
    \ket{\mu}\bra{\mu} &=& \begin{pmatrix} \mu & e^{-i\alpha_\mu}\sqrt{\mu(1-\mu)} \\ e^{i\alpha_\mu}\sqrt{\mu(1-\mu)} & 1- \mu \end{pmatrix}, \\
    \ket{\nu}\bra{\nu} &=& \begin{pmatrix} \nu & e^{-i\alpha_\nu}\sqrt{\nu(1-\nu)} \\ e^{i\alpha_\nu}\sqrt{\nu(1-\nu)} & 1- \nu \end{pmatrix}.    
\end{eqnarray}
Hence, the condition of local thermality $p |\mu \rangle \langle \mu | + (1-p) |\nu \rangle \langle \nu | = \tau_{\beta_{\rm in}}$ can be provided in the form of two equations that define $p$, $\nu$, and $\alpha_\nu$:
\begin{eqnarray}
    p\mu + (1-p)\nu &=& q, \\
    e^{\pm i\alpha_\mu} p \sqrt{\mu (1-\mu)} + e^{\pm i\alpha_\nu} (1-p) \sqrt{\nu (1-\nu)} &=& 0,
\end{eqnarray}
where $q = \frac{1}{Z_{S, \beta_{\rm in}}}$, i.e., defined by the equation $\beta_{\rm in} = \ln\frac{q}{1-q}$. Solving them we find $p = \frac{q(1-q)}{q + \mu - 2\mu q}$, $\nu = \frac{2q^2(1-\mu)}{q^2 + \mu(1 - 2q)}$, and $\alpha_\nu = \alpha_\mu + \pi k$, $k \in \mathbb{Z}$. Plugging them in into~(\ref{app:eq:PureDiscord}), we obtain the form~(\ref{app:eq:locthermoptD}) of the optimal initial state, hence the proof.
    \end{proof}
\end{lemma}

Finally, we provide the calculation of the resulting maximal SWITCH-ergotropy, as shown in the following Proposition.

\begin{proposition}
    The SWITCH-ergotropy~(\ref{def:SwitchErg}) with joint input state with quantum discord is maximal for
            \begin{equation}
    \rho_{SC}^{\mathrm{opt}} = \left\{
 \begin{array}{lc}
 \frac{1}{Z_{S,\beta_{\rm in}}} \Bigl( \ket{0}\bra{0} \otimes \ket{+}\bra{+} + e^{-\beta_{\rm in}} \ket{1}\bra{1}\otimes \ket{-}\bra{-}\Bigr), & \text{for} ~\beta_{\rm in} \geq 2\beta  \\
 \frac12 \Bigl( \ket{\mu}\bra{\mu} \otimes \ket{+}\bra{+} + \ket{\nu}\bra{\nu} \otimes \ket{-}\bra{-} \Bigr), & \text{for} ~\beta_{\rm in} \leq 2\beta 
\end{array}\right. ,
    \end{equation}
where $|\mu/\nu\rangle = \frac{1}{\sqrt{Z_{S, \beta_{\rm in}}}} \Bigl( |0\rangle \pm e^{-\frac{\beta_{\rm in}}{2}} |1\rangle \Bigr)$, and is given by:
\begin{equation}
    \mathcal{W}_{\mathrm{D}} = \left\{
 \begin{array}{lc}
 \frac{1}{2}\tanh\Bigl(\frac{\beta}{2}\Bigr)\Biggl( \frac{\cosh\Bigl(\beta + \frac{\beta_{\rm in}}{2}\Bigr)}{2\sinh(\beta)\cosh\Bigl(\frac{\beta_{\rm in}}{2}\Bigr)} - 1\Biggr), & \text{for} ~\beta_{\rm in} \geq 2\beta  \\
 \frac{1}{2}\tanh\Bigl(\frac{\beta}{2}\Bigr)\Biggl( \sqrt{1 + \frac{1}{4}\sinh^{-2}(\beta)\cosh^{-2}\Bigl(\frac{\beta_{\rm in}}{2}\Bigr)} - 1\Biggr), & \text{for} ~\beta_{\rm in} \leq 2\beta 
\end{array}\right. ,
    \end{equation}
    $\mathcal{W}_{\mathrm{D}}$ is purely incoherent for $\beta_{\rm in} \geq 2\beta$ and purely coherent for $\beta_{\rm in} \leq 2\beta$.
 \begin{proof}        

Applying Lemma~\ref{app:lemma:locthermoptD}, we can write the maximal SWITCH-ergotropy~(\ref{app:eq:OptErgD}) in a simpler form:
\begin{align}
    \mathcal{W}_{\rm D} &= \frac{1}{2}\max_\mu \Biggl( W\Bigl( \tau_\beta + (G_{\rm CC} + G_{\rm coh}) \Bigr) + W\Bigl( \tau_\beta - (G_{\rm CC} + G_{\rm coh}) \Bigr)\Biggr), \\
    G_\mathrm{CC} & = \frac{\mu-q}{\mu+q-2q\mu}\tau_\beta \sqrt{\tau_{\beta_\mathrm{in}}} Z \sqrt{\tau_{\beta_\mathrm{in}}} \tau_\beta, \\
    G_{\rm coh} &= \frac{2q(1-q) \sqrt{\mu(1-\mu)}}{\mu+q-2q\mu} \tau_\beta X \tau_\beta,
\end{align}
where $\frac{1}{2} \leq q \leq 1$ is defined by $\beta_{\rm in} = \ln\frac{q}{1-q}$. In turn, we can separate the SWITCH-ergotropy into its incoherent and coherent counterparts and calculate them explicitly using~(\ref{eq:IncohErg}) and~(\ref{eq:CohErg}), respectively:
\begin{align}
    \mathcal{W}_{\rm D} &= \max_\mu \Bigl( W_{\rm inc}^{\rm D}(\mu) + W_{\rm coh}^{\rm D}(\mu) \Bigr), \\
    W_{\rm inc}^{\rm D}(\mu) &= \frac{1}{2} \tanh\Bigl(\frac{\beta}{2}\Bigr) \max\Biggl\{ 0, \frac{|q-\mu|}{\mu+q-2q\mu}\frac{\cosh\Bigl(\beta + \frac{\beta_{\rm in}}{2}\Bigr)}{2\sinh(\beta)\cosh\Bigl(\frac{\beta_{\rm in}}{2}\Bigr)} - 1 \Biggr\}, \\
    \nonumber W_{\rm coh}^{\rm D}(\mu) &= 
    \frac{1}{4} \tanh\Bigl(\frac{\beta}{2}\Bigr) \Biggl[ \sqrt{(1 + \Xi)^2 + \frac{4q(1-q)\mu(1-\mu)}{(\mu+q-2q\mu)^2} \Biggl( \frac{1}{2\sinh(\beta) \cosh\Bigl(\frac{\beta_{\rm in}}{2}\Bigr)} \Biggr)^2} \\
    &+ \sqrt{(1 - \Xi)^2 + \frac{4q(1-q)\mu(1-\mu)}{(\mu+q-2q\mu)^2} \Biggl( \frac{1}{2\sinh(\beta) \cosh\Bigl(\frac{\beta_{\rm in}}{2}\Bigr)} \Biggr)^2} - | 1 + \Xi| - | 1 - \Xi| \Biggr],
\end{align}
where
\begin{align}
    \Xi &= \frac{q-\mu}{\mu+q-2q\mu} \frac{\cosh\Bigl(\beta + \frac{\beta_{\rm in}}{2}\Bigr)}{2\sinh(\beta) \cosh\Bigl(\frac{\beta_{\rm in}}{2}\Bigr)}.
\end{align}
Performing optimization of the overall SWITCH-ergotropy, we find two regimes of temperature pairs $(\beta, \beta_{\rm in})$, which feature different optimal initial states $\rho_{SC}$ and are separated by the temperature bound~(\ref{eq:TempLimit}). If $\beta_{\rm in} \geq 2\beta$, the optimal value $\mu = 1$ defines the initial state
\begin{eqnarray}\label{app:DiscordMaxIncoh}
   \rho_{SC}^{\mathrm{opt}, \beta_{\rm in} \geq 2\beta} = \frac{1}{Z_{S,\beta_{\rm in}}} \Bigl( \ket{0}\bra{0} \otimes \ket{+}\bra{+} + e^{-\beta_{\rm in}} \ket{1}\bra{1}\otimes \ket{-}\bra{-}\Bigr),
\end{eqnarray}
and the corresponding maximal SWITCH-ergotropy coincides with the maximal SWITCH-ergotropy under classical correlations guaranteed by Theorem~\ref{theor:ErgCC}, hence, containing only the incoherent counterpart:
\begin{equation}\label{app:DiscordMaxErgInc}
\mathcal{W}_{\rm D}^{\beta_{\rm in} \geq 2\beta} = \frac{1}{2Z_\beta^2Z_{\beta_{\rm in}}} \max\{0, e^{-2\beta} - e^{-\beta_{\rm in}} + 2e^{-(2\beta+\beta_{\rm in})} \}.
\end{equation}
On the other hand, if the temperature bound~(\ref{eq:TempLimit}) is violated, i.e., $\beta_{\rm in} \leq 2\beta$, the optimal initial state is defined by $\mu = q \equiv \frac{1}{Z_{\beta_{\rm in}}}$, so that:
\begin{eqnarray}\label{app:DiscordMaxCoh}
   \rho_{SC}^{\mathrm{opt}, \beta_{\rm in} \leq 2\beta} = \frac12 \Bigl[ \ket{\mu}\bra{\mu} \otimes \ket{+}\bra{+} + \ket{\nu}\bra{\nu} \otimes \ket{-}\bra{-} \Bigr],
\end{eqnarray}
with $|\mu/\nu\rangle = \frac{1}{\sqrt{Z_{S, \beta_{\rm in}}}} \Bigl(|0\rangle \pm e^{-\frac{\beta_{\rm in}}{2}} |1\rangle\Bigr)$. In this case, the resulting maximal SWITCH-ergotropy consists of the coherent counterpart only,
\begin{equation}\label{app:DiscordMaxErgCoh}
\mathcal{W}_{\rm D}^{\beta_{\rm in} \leq 2\beta} = \frac{1}{2}\tanh\Bigl(\frac{\beta}{2}\Bigr)\Biggl( \sqrt{1 + \frac{1}{4}\sinh^{-2}(\beta)\cosh^{-2}\Bigl(\frac{\beta_{\rm in}}{2}\Bigr)} - 1\Biggr),
\end{equation}
which coincides with the maximal SWITCH-ergotropy~(\ref{app:eq:ErgCC:zero}) under classical correlations if $\beta_{\rm in} = 0$. Hence the proof.
    \end{proof}
\end{proposition}

\section{Proof of Theorem~\ref{theor:opt_ent}}
\label{appQC}

In this Appendix, we find the maximal SWITCH-ergotropy that can be achieved under entangled initial joint state of work medium (under the condition of local thermality) and the control system, 
\begin{align}\label{app:eq:EntState}
    \rho_{SC} &= |\Phi_{\alpha, \phi}(\beta_{\rm in})\rangle\langle\Phi_{\alpha, \phi}(\beta_{\rm in})|, \\
    |\Phi_{\alpha, \phi}(\beta_{\rm in})\rangle &= \frac{1}{\sqrt{Z_{S,\beta_{\rm in}}}} \Bigl( |0\psi_{\alpha, \phi}\rangle + e^{-\frac{\beta_{\rm in}}{2}}|1\psi_{\alpha, \phi}^{\perp}\rangle \Bigr),
\end{align}

\begin{lemma}\label{app:lemma:OutputQC}
    Given a purification $(\alpha, \phi)$ of the initial thermal state $\tau_{\beta_{\rm in}}$ of the work medium with respect to the control system, the quantum SWITCH of two thermalizing channels $\mathcal{T}_\beta[\cdot]$ outputs a joint state
    \begin{align}\label{app:eq:OutputQC}
        \rho_{\rm ENT} = \begin{bmatrix}
        p (1 {-} \alpha {+} q (-1 {+} 2 \alpha)) & e^{-i \phi} p^2 q \sqrt{\alpha(1 {-} \alpha)} & 0 & -(1 {-} p) p \alpha \sqrt{q (1{-}q)} \\
        e^{i \phi} p^2q\sqrt{\alpha(1{-}\alpha)} & p (\alpha {+}  q (1 {-} 2 \alpha)) & e^{2 i \phi}(1 {-} p) p (1{-}\alpha) \sqrt{q (1{-}q)} & 0 \\
        0 & e^{-2 i \phi}(1 {-} p) p (1{-}\alpha) \sqrt{q (1{-}q)} & (1 {-} p) (1 {-} \alpha {+} q (-1 {+} 2 \alpha)) & -e^{-i \phi}(1 {-} p)^2 (1{-}q) \sqrt{\alpha (1 {-} \alpha)} \\
        -(1 {-} p) p \alpha \sqrt{q (1{-}q)} & 0 & -e^{i \phi}(1 {-} p)^2 (1{-}q) \sqrt{\alpha (1 {-} \alpha)} & (1 {-} p) (\alpha {+} q (1 {-} 2 \alpha))
    \end{bmatrix},
    \end{align}
    where the computational basis $\{\ket{i} \otimes \ket{j}\}$ with $i,j \in \{0,1\}$ is used, while $p = \frac{1}{Z_{S, \beta}} \in [\frac{1}{2}, 1]$ and $q = \frac{1}{Z_{S, \beta_{\rm in}}} \in [\frac{1}{2}, 1]$.
\begin{proof}
We start by taking the purification $(\alpha, \phi)$ with respect to~(\ref{eq:purification}) and rewriting it as
\begin{equation}
|\Phi_{\alpha, \phi}(\beta_{\rm in})\rangle = \frac{1}{\sqrt{Z_{S,\beta_{\rm in}}}} \Bigl( \sqrt{\alpha} |00\rangle + e^{i \phi} \sqrt{1-\alpha}|01\rangle + e^{-\frac{\beta_{\rm in}}{2}}e^{-i \phi} \sqrt{1-\alpha}|10\rangle - e^{-\frac{\beta_{\rm in}}{2}} \sqrt{\alpha}|11\rangle \Bigr).
\end{equation}
In turn, plugging $\rho_{SC} = |\Phi_{\alpha, \phi}(\beta_{\rm in})\rangle\langle\Phi_{\alpha, \phi}(\beta_{\rm in})|$ in into the quantum SWITCH channel~(\ref{eq:switchchannel}) we obtain
\begin{align}
    \nonumber \rho_{\rm ENT} &= \frac{1}{Z_{S,\beta_{\rm in}}} \sum_{iji'j'} \Bigl[ T_{\beta,ij} T_{\beta,i'j'} \Bigl(\alpha |0\rangle\langle 0| + \sqrt{\alpha(1-\alpha)} e^{-\frac{\beta_{\rm in}}{2}} (e^{i\phi}|0\rangle \langle 1 | + e^{-i\phi}|1\rangle \langle 0 |) \\
    \nonumber &+ e^{-\beta_{\rm in}} (1-\alpha) |1\rangle\langle 1| \Bigr) (T_{\beta,ij} T_{\beta,i'j'})^\dagger \otimes |0\rangle \langle 0 | \\
\nonumber &+ T_{\beta,i'j'} T_{\beta,ij} \Bigl((1-\alpha) |0\rangle\langle 0| + \sqrt{\alpha(1-\alpha)} e^{-\frac{\beta_{\rm in}}{2}} (e^{i\phi}|0\rangle \langle 1 | + e^{-i\phi}|1\rangle \langle 0 |) \\
\nonumber &+ e^{-\beta_{\rm in}} \alpha |1\rangle\langle 1| \Bigr) (T_{\beta,i'j'} T_{\beta,ij} )^\dagger \otimes |1\rangle \langle 1 | \\
\nonumber &+ T_{\beta,ij} T_{\beta,i'j'} \Bigl( e^{-i \phi} \sqrt{\alpha(1-\alpha)} \Bigl( |0\rangle\langle 0| - e^{-\beta_{\rm in}} |1\rangle\langle 1| \Bigr) - e^{-\frac{\beta_{\rm in}}{2}} \alpha |0\rangle\langle 1| \\
\nonumber &+ e^{-\frac{\beta_{\rm in}}{2}} e^{-2i \phi} (1-\alpha) |1\rangle\langle 0| \Bigr) (T_{\beta,i'j'} T_{\beta,ij} )^\dagger \otimes |0\rangle \langle 1 | \\
\nonumber &+ T_{\beta,i'j'} T_{\beta,ij} \Bigl( e^{i \phi} \sqrt{\alpha(1-\alpha)}\Bigl( |0\rangle\langle 0| - e^{-\beta_{\rm in}} |1\rangle\langle 1| \Bigr) + e^{-\frac{\beta_{\rm in}}{2}} e^{2i \phi} (1-\alpha) |0\rangle\langle 1| \\
&- e^{-\frac{\beta_{\rm in}}{2}} \alpha |1\rangle\langle 0| \Bigr) (T_{\beta,ij} T_{\beta,i'j'} )^\dagger \otimes |1\rangle \langle 0 | \Bigr],
\end{align}
where $\rho_{\rm ENT} = \big[ \map S (\mathcal{T}_\beta, \mathcal{T}_\beta) \big]  (\ket{\Phi_{\alpha, \phi}(\beta_{\rm in})}\bra{\Phi_{\alpha, \phi}(\beta_{\rm in})})\Big)$ is the output of the quantum SWITCH and $\{T_{\beta,ij}\}_{i,j}$ are the Kraus operators of $\mathcal{T}_\beta$ defined in~(\ref{eq:ThermKraus}). Calculating explicitly the action of the Kraus operators, we obtain
\begin{align}
\nonumber \rho_{\rm ENT} &= \frac{1}{Z_{S,\beta_{\rm in}}} \Bigl[  \Bigl(\alpha \tau_\beta + e^{-\beta_{\rm in}} (1-\alpha) \tau_\beta \Bigr) \otimes |0\rangle \langle 0 | + \Bigl((1-\alpha) \tau_\beta + e^{-\beta_{\rm in}} \alpha \tau_\beta \Bigr) \otimes |1\rangle \langle 1 | \\
\nonumber &+ \sqrt{\alpha(1-\alpha)}  \tau_\beta \Bigl( |0\rangle\langle 0| - e^{-\beta_{\rm in}} |1\rangle\langle 1| \Bigr) \tau_\beta \otimes  \Bigl( e^{-i \phi} |0\rangle \langle 1 | + e^{i \phi} \otimes |1\rangle \langle 0 | \Bigr) \\
\nonumber &+ \tau_\beta\Bigl( - e^{-\frac{\beta_{\rm in}}{2}} \alpha |0\rangle\langle 1| + e^{-\frac{\beta_{\rm in}}{2}} e^{-2i \phi} (1-\alpha) |1\rangle\langle 0| \Bigr)\tau_\beta \otimes |0\rangle \langle 1 | \\
&+ \tau_\beta\Bigl( e^{-\frac{\beta_{\rm in}}{2}} e^{2i \phi} (1-\alpha) |0\rangle\langle 1| - e^{-\frac{\beta_{\rm in}}{2}} \alpha |1\rangle\langle 0| \Bigr)\tau_\beta \otimes |1\rangle \langle 0 |\Bigr].
\end{align}
Finally, writing the thermal state $\tau_\beta$ in its matrix form and redefining $\beta = \ln{\frac{p}{1-p}}$ and $\beta_{\rm in} = \ln{\frac{q}{1-q}}$, we obtain~(\ref{app:eq:OutputQC}), hence the proof.
\end{proof}
\end{lemma}

Finally, we provide the calculation of the resulting maximal SWITCH-ergotropy, as shown in the following Proposition.

\begin{proposition}
The SWITCH-ergotropy~(\ref{def:SwitchErg}) with entangled joint input state is maximal for purification with
\begin{eqnarray}
\alpha_{\rm{opt}}(\beta_{\rm in}, \beta) &=&  \left\{
 \begin{array}{cc}
 0 \text{ or } 1, & \text{for} ~\beta_{\rm in} \leq 2 \beta  \\
 \frac12, & \text{for} ~\beta_{\rm in} \geq 2 \beta
\end{array}\right., 
\end{eqnarray}
regardless of $\phi$ and is given by:
\begin{equation}
    \mathcal{W}_{\mathrm{ENT}} = \left\{
 \begin{array}{lc}
 \frac{1}{2}\tanh\Bigl(\frac{\beta}{2}\Bigr)\Biggl( \frac{\cosh\Bigl(\beta + \frac{\beta_{\rm in}}{2}\Bigr)}{2\sinh(\beta)\cosh\Bigl(\frac{\beta_{\rm in}}{2}\Bigr)} - 1\Biggr), & \text{for} ~\beta_{\rm in} \geq 2\beta  \\
 \frac{1}{2}\tanh\Bigl(\frac{\beta}{2}\Bigr)\Biggl( \sqrt{1 + \frac{1}{4\sinh(\beta)^2\cosh^2\Bigl(\frac{\beta_{\rm in}}{2}\Bigr)}} - 1\Biggr), & \text{for} ~\beta_{\rm in} \leq 2\beta 
\end{array}\right. ,
    \end{equation}

\begin{proof}
Applying Lemma~\ref{app:lemma:OutputQC}, we rewrite the output state $\rho_{\rm ENT}$ in a simpler form:
\begin{align}
    \rho_{\rm ENT} &= \frac{1}{2} \Bigl[ \tau_\beta \otimes (\mathds{1} + (1-2\alpha)(1-2q) Z) + G_X \otimes X + G_Y \otimes Y \Bigr], \\
    G_X &= \tau_\beta \sqrt{\tau_{\beta_{\rm in}}} \begin{pmatrix} 2 \cos\phi \sqrt{\alpha(1-\alpha)} & (1-\alpha) e^{2i\phi} - \alpha \\ (1-\alpha) e^{-2i\phi} - \alpha & -2 \cos\phi \sqrt{\alpha(1-\alpha)} \end{pmatrix} \sqrt{\tau_{\beta_{\rm in}}} \tau_\beta, \\
    G_Y &= \tau_\beta \sqrt{\tau_{\beta_{\rm in}}} \begin{pmatrix} 2 \sin\phi \sqrt{\alpha(1-\alpha)} & -i((1-\alpha) e^{2i\phi} + \alpha) \\ i((1-\alpha) e^{-2i\phi} + \alpha) & -2 \sin\phi \sqrt{\alpha(1-\alpha)} \end{pmatrix} \sqrt{\tau_{\beta_{\rm in}}} \tau_\beta.
\end{align}
Therefore, the corresponding maximal SWITCH-ergotropy can be written as:
\begin{align}
    \nonumber \mathcal{W}_{\rm ENT} &= \frac{1}{2} \max_{\alpha, \phi, |\psi\rangle} \Biggl( W\Bigl( (1 + \tilde{p}(\psi, \alpha)) \tau_\beta + \langle X \rangle_\psi G_X + \langle Y \rangle_\psi G_Y  \Bigr) \\
    &+ W\Bigl( (1 - \tilde{p}(\psi, \alpha)) \tau_\beta - \langle X \rangle_\psi G_X - \langle Y \rangle_\psi G_Y  \Bigr) \Biggr),
\end{align}
where $\tilde{p}(\psi, \alpha) = (1-2\alpha)(1-2q) \langle Z \rangle_\psi$. The terms proportional to $G_X$ and $G_Y$ are the only ones contributing to the SWITCH-ergotropy. Hence, its optimization requires maximization of these terms. Expanding $|\psi\rangle = \begin{pmatrix} \cos\Bigl(\frac{\theta'}{2}\Bigr) \\  e^{i\phi'}\sin\Bigl(\frac{\theta'}{2}\Bigr)\end{pmatrix}$, we provide the components of its Bloch vector as $\langle X \rangle_\psi = \sin\theta' \cos\phi'$ and $\langle Y \rangle_\psi = \sin\theta' \cos\phi'$. This leads to:
\begin{equation}
    \langle X \rangle_\psi G_X + \langle Y \rangle_\psi G_Y = \tau_\beta \sqrt{\tau_{\beta_{\rm in}}} \begin{pmatrix} 2 \cos(\phi-\phi') \sin(\theta') \sqrt{\alpha(1-\alpha)} & e^{-i\phi'}((1-\alpha) e^{2i(\phi-\phi')} - \alpha) \\ e^{-i\phi'}((1-\alpha) e^{-2i(\phi-\phi')} - \alpha) & -2 \cos(\phi-\phi') \sin(\theta') \sqrt{\alpha(1-\alpha)} \end{pmatrix}  \sqrt{\tau_{\beta_{\rm in}}} \tau_\beta.
\end{equation}
Therefore, its contribution is maximal if the measurement is performed with respect to the basis $\{|\psi\rangle, |\psi^\perp\rangle\}$ with $\theta' = \frac{\pi}{2} + \pi k$ and $\phi' = \phi + \pi k'$, where $k, k' \in \mathbb{Z}$. We obtain hence the SWITCH-ergotropy:
\begin{align}\label{app:eq:EntErg}
    \mathcal{W}_{\rm ENT} &= \frac{1}{2} \max_{\alpha, \phi} \Biggl( W\Bigl( \tau_\beta + G \Bigr) + W\Bigl( \tau_\beta - G  \Bigr) \Biggr), \\
    G &= \tau_\beta \sqrt{\tau_{\beta_{\rm in}}} \begin{pmatrix} 2 \sqrt{\alpha(1-\alpha)} & e^{i\phi}(1- 2\alpha) \\ e^{-i\phi}(1-2\alpha) & -2 \sqrt{\alpha(1-\alpha)} \end{pmatrix} \sqrt{\tau_{\beta_{\rm in}}} \tau_\beta.
\end{align}
Now, we can calculate separately the incoherent and coherent counterparts of~(\ref{app:eq:EntErg}) by applying~(\ref{eq:IncohErg}) and~(\ref{eq:CohErg}), respectively:
\begin{align}
    \mathcal{W}_{\rm ENT} &= \max_\alpha \Bigl( W_{\rm inc}^{\rm ENT}(\alpha) + W_{\rm coh}^{\rm ENT}(\alpha) \Bigr), \\
    W_{\rm inc}^{\rm ENT}(\alpha) &= \frac{1}{2} \tanh\Bigl(\frac{\beta}{2}\Bigr) \max\Biggl\{ 0, \sqrt{\alpha(1-\alpha)} \frac{\cosh\Bigl(\beta + \frac{\beta_{\rm in}}{2}\Bigr)}{\sinh(\beta) \cosh\Bigl(\frac{\beta_{\rm in}}{2}\Bigr)} - 1 \Biggr\}, \\
    \nonumber W_{\rm coh}^{\rm ENT}(\alpha) &= \frac{1}{4} \tanh\Bigl(\frac{\beta}{2}\Bigr) \Biggl[ \sqrt{\Biggl(1 + \sqrt{\alpha(1-\alpha)} \frac{\cosh\Bigl(\beta + \frac{\beta_{\rm in}}{2}\Bigr)}{\sinh(\beta) \cosh\Bigl(\frac{\beta_{\rm in}}{2}\Bigr)}\Biggr)^2 + \Biggl( \frac{1-2\alpha}{2\sinh(\beta) \cosh\Bigl(\frac{\beta_{\rm in}}{2}\Bigr)} \Biggr)^2} \\
    \nonumber &+ \sqrt{\Biggl(1 - \sqrt{\alpha(1-\alpha)} \frac{\cosh\Bigl(\beta + \frac{\beta_{\rm in}}{2}\Bigr)}{\sinh(\beta) \cosh\Bigl(\frac{\beta_{\rm in}}{2}\Bigr)}\Biggr)^2 + \Biggl( \frac{1-2\alpha}{2\sinh(\beta) \cosh\Bigl(\frac{\beta_{\rm in}}{2}\Bigr)} \Biggr)^2} \\
    &- \Biggl| 1 + \sqrt{\alpha(1-\alpha)} \frac{\cosh\Bigl(\beta + \frac{\beta_{\rm in}}{2}\Bigr)}{\sinh(\beta) \cosh\Bigl(\frac{\beta_{\rm in}}{2}\Bigr)} \Biggr| - \Biggl| 1 - \sqrt{\alpha(1-\alpha)} \frac{\cosh\Bigl(\beta + \frac{\beta_{\rm in}}{2}\Bigr)}{\sinh(\beta) \cosh\Bigl(\frac{\beta_{\rm in}}{2}\Bigr)} \Biggr| \Biggr].
\end{align}
While the overall maximal SWITCH-ergotropy is invariant under change of $\phi$, its optimization over $\alpha$ reveals two regimes of temperature pairs $(\beta, \beta_{\rm in})$ featuring different optimal values of $\alpha$ and separated by the temperature bound~(\ref{eq:TempLimit}). If $\beta_{\rm in} \geq 2\beta$, the optimal value $\alpha = \frac{1}{2}$ provides the maximal SWITCH-ergotropy of exclusively incoherent nature given by:
\begin{equation}
    \mathcal{W}_{\rm ENT}^{\beta_{\rm in} \geq 2\beta} = \frac{1}{2}\tanh\Bigl(\frac{\beta}{2}\Bigr)\Biggl( \frac{\cosh\Bigl(\beta + \frac{\beta_{\rm in}}{2}\Bigr)}{2\sinh(\beta)\cosh\Bigl(\frac{\beta_{\rm in}}{2}\Bigr)} - 1\Biggr).
\end{equation}
On the other hand, if the temperature bound~(\ref{eq:TempLimit}) is violated, i.e., $\beta_{\rm in} \leq 2\beta$, the optimal purification is defined by $\alpha = 0$ or $\alpha = 1$, so that the resulting maximal SWITCH-ergotropy consists of its coherent counterpart only:
\begin{equation}
    \mathcal{W}_{\rm ENT}^{\beta_{\rm in} \leq 2\beta} = \frac{1}{2}\tanh\Bigl(\frac{\beta}{2}\Bigr)\Biggl( \sqrt{1 + \frac{1}{4\sinh(\beta)^2\cosh^2\Bigl(\frac{\beta_{\rm in}}{2}\Bigr)}} - 1\Biggr),
\end{equation}
hence the proof.
\end{proof}
\end{proposition}

\section{Initial mixed entangled state of $S$ and $C$}\label{app:mixedQC}
As a case study, let us consider a one-parameter family of mixed entangled states
\begin{eqnarray}
    \chi_{\lambda} = \lambda |\Phi_{\alpha,\phi}(\beta_{\text{in}})\rangle \langle \Phi_{\alpha,\phi}(\beta_{\text{in}})| + (1-\lambda)\tau_{\beta_{\text{in}}} \otimes \frac{\mathds{1}_C}{2}.
\end{eqnarray}
Since the action of the quantum SWITCH is linear due to \eqref{eq:switchchannel}, its output is given by:
\begin{equation}
    \big[\mathcal{S}(\mathcal{T}_{\beta}, \mathcal{T}_{\beta})\big] (\chi_\lambda) = \lambda \rho_{\mathrm{ENT}} + (1-\lambda)\tau_{\beta} \otimes \frac{\mathds{1}_C}{2},
\end{equation}
where $\rho_{\mathrm{ENT}} = \big[\mathcal{S}(\mathcal{T}_{\beta}, \mathcal{T}_{\beta})\big](|\Phi_{\alpha,\phi}(\beta_{\text{in}})\rangle \langle \Phi_{\alpha,\phi}(\beta_{\text{in}})|)$ is the output of the quantum SWITCH for an initial pure entangled state of $S$ and $C$ with a thermal marginal and is given by \eqref{app:eq:OutputQC}. On the other hand, sublinearity of ergotropy \cite{Bernards2019} suggests that
\begin{equation}\label{app:eq:sublinErg}
    W_{\mathrm{dae}}\Bigl(\big[\mathcal{S}(\mathcal{T}_{\beta}, \mathcal{T}_{\beta})\big] (\chi_\lambda)\Bigr) \leq \lambda W_{\mathrm{dae}}(\rho_{\mathrm{ENT}}) + (1-\lambda)W_{\mathrm{dae}}\Bigl(\tau_{\beta} \otimes \frac{\mathds{1}_C}{2}\Bigr),
\end{equation}
As highlighted in Section \ref{sec:thermalStates}, thermal states are completely passive and, hence, carry zero ergotropy. Therefore, the last term in \eqref{app:eq:sublinErg} is zero, and we obtain
\begin{equation}
    W_{\mathrm{dae}}\Bigl(\big[\mathcal{S}(\mathcal{T}_{\beta}, \mathcal{T}_{\beta})\big] (\chi_\lambda)\Bigr) \leq \lambda W_{\mathrm{dae}}(\rho_{\mathrm{ENT}}),
\end{equation}
suggesting that the optimal ergotropy is achieved by $\rho_{\mathrm{ENT}}$ corresponding to initial purified thermal state.

\end{document}